\documentclass[a4paper,12pt]{article}
\pdfoutput=1 

\usepackage{jheppub} 
\usepackage[normalem]{ulem}
\usepackage[T1]{fontenc} 
\usepackage{xcolor}
\usepackage{mathrsfs} 
\usepackage{subfigure}

\title{\boldmath A Study of the SYK$_{2}$ Model with Twisted Boundary Conditions}

\author[a,b]{Jeff Murugan,}
\author[a]{Ruach Pillay Slayen}
\author[a,b]{and Hendrik J. R. Van Zyl}

\affiliation[a]{The Laboratory for Quantum Gravity \& Strings, Department of Mathematics \& Applied Mathematics, University of Cape Town, Cape Town, South Africa}
\affiliation[b]{National Institute for Theoretical and Computational Sciences (NITheCS),\\ South Africa}

\dedicated{Dedicated to the memory of Roman Jackiw: A pioneer of many dimensions, in few.}

\abstract{We study a version of the 2-body Sachdev-Ye-Kitaev (SYK$_{2}$) model whose complex fermions exhibit twisted boundary conditions on the thermal circle. As we show, this is physically equivalent to coupling the fermions to a 1-dimensional external gauge field $A(t)$. In the latter formulation, the gauge field itself can be thought of as arising from a radial symmetry reduction of a $(2+1)$-dimensional Chern-Simons gauge field $A_{\mu}(t,\mathbf{x})$. Using the diagnostic tools of the out-of-time-order correlator (OTOC) and spectral form factor (SFF), which probe the sensitivity to initial conditions and the spectral statistics respectively, we give a detailed and pedagogical study of the integrable/chaotic properties of the model. We find that the twisting has no effect on the OTOCs and, by extension, the early-time chaos properties of the model. It does, however, have two notable effects on the spectral form factor; an enhancement of the early-time slope and the emergence of an explicit disorder scale needed for the manifestation of zero modes. These zero modes are responsible for the late-time exponential ramp in the quadratic SYK model.}

\begin{document} 
\maketitle
\section{Introduction}\label{intro}
Sachdev-Ye-Kitaev (SYK) models \cite{Sachdev:1992fk, Maldacena:2016hyu, Polchinski:2016xgd, Garcia-Garcia:2016mno, Gross:2016kjj, Gross:2017aos, Saad:2018bqo, Rosenhaus:2018dtp, Gu:2019jub} detail the properties of a family of (0+1)-dimensional quantum mechanics of $N\gg1$ Majorana fermions, $\chi_{a}$, with all-to-all random $q$-fermi interactions. The $q=0$ free theory with action,
\begin{eqnarray}
   I_{\mathrm{free}} = \sum_{a=1}^{N}\int\!\!dt\,\,\chi_{a}\dot{\chi}_{a}\,,
   \label{free}
\end{eqnarray} 
is an exactly conformal quantum mechanics of $2^{N/2}$ zero-energy states and a global $SO(N)$ flavour symmetry. The free theory admits $N$ relevant deformations implemented by perturbing \eqref{free} with fermion monomial operators. In this sense, the standard ($q=4$) SYK model \cite{Sachdev:1992fk}, 
\begin{eqnarray}
   I_{\mathrm{SYK}} = \int\!\!dt\left(\sum_{a=1}^{N}\,\,\chi_{a}\dot{\chi}_{a} + \sum_{a,b,c,d=1}^{N}
   \frac{1}{4!}J_{abcd}\,\,\chi_{a}\chi_{b}\chi_{c}\chi_{d}\right)\,,
   \label{SYK}
\end{eqnarray} 
can be considered a perturbation of \eqref{free} by a quartic monomial that preserves the full $SO(N)$ flavour group as long as the fermion coupling $J_{abcd}$ is disordered. Here the couplings are time-independent (and so the disorder is quenched) and drawn randomly from a Gaussian distribution with $\overline{J_{abcd}} = 0$ and $\overline{J_{abcd}J^{abcd}} = 3!J^{2}/N^{3}$.  This model has some well-documented and remarkable properties \cite{Maldacena:2016hyu}:
\begin{itemize}
   \item It is {\it solvable}. At zero temperature and large $N$, the solution to the Schwinger-Dyson equation for the fermion 2-point function is
   \begin{eqnarray}
      \langle \chi_{a}(t)\chi_{b}(0) \rangle  = \left(\frac{1}{4\pi J^{2}}\right)^{1/4}\frac{\mathrm{sgn(t)}}{|t|^{1/2}}\delta_{ab}\,,
   \end{eqnarray}
   in the infrared and for $t\gg 1/J$. Dimensional analysis of this solution shows that $[\chi] = 1/4$ in the IR.
   \item Indeed, the SYK model manifests an {\it emergent conformal symmetry} at low energies and large $N$.
   \item The model is famously {\it maximally chaotic} in the sense that the quantum Lyapunov exponent, extracted from the out-of-time-ordered thermal fermion 4-point function, saturates the Maldacena-Shenker-Stanford (MSS) bound \cite{Maldacena:2015waa} $\lambda_{L}\leq 2\pi T$ at temperature $T$.
\end{itemize} 
With all this going for it, one might very reasonably presume the SYK model to furnish the CFT dual to an AdS${}_{2}$/CFT${}_{1}$ correspondence \cite{Sarosi:2017ykf, Almheiri:2014cka, Jevicki:2016bwu, Engelsoy:2016xyb, Jevicki:2016ito, Cvetic:2016eiv}. However, this hope was convincingly dashed in \cite{Jensen:2016pah} where it was pointed out that the SYK model {\it cannot} have a conventional weakly-coupled, weakly-curved gravity dual for two crucial reasons:
\begin{enumerate}
   \item The large $N$ Majoranas $\chi_{a}(t)$ of the SYK model are dual to $N$ degenerate bulk fermions $\Psi_{a}(t,x)$ whose contribution to the bulk partition function is comparable to the classical saddle, consequently invalidating the saddle-point approximation.
   \item The model does not exhibit the large $N$ factorization expected of theories with a conventional gravitational dual.
\end{enumerate}
Interpreting these observations is tricky. In \cite{Jensen:2016pah} for example, it was suggested that perhaps the correct statement is that the singlet sector of the SYK model is dual to a 2-dimensional higher spin theory. To support this assertion, it was shown there that the gravitational dynamics near an AdS${}_{2}$ throat can be re-written in terms of an effective quantum hydrodynamical action that precisely matches with the Schwarzian action found in \cite{Maldacena:2016hyu} from the low energy, large $N$ solution of the SYK model. Another view point is that the ills of the SYK model could be cured by gauging a large subgroup of the $SO(N)$ global flavour symmetry \cite{Jensen:2016pah}. This has, however, proven a difficult goal to realize.\\

\noindent
With the goal of constructing such an AdS$_{2}$/gauged-SYK correspondence in mind then, we began a study of a generalization of the original SYK model, in which the $N$ Majoranas are replaced by $N\gg1$ spinless complex fermions - the so-called {\it charged} SYK model \cite{Sachdev:2015efa,Bulycheva:2017uqj}. This system possesses a global $U(1)$ symmetry and for the case $q=4$, for example, is described by the action 
\begin{eqnarray}
   I_{\mathrm{CSYK}} = \int dt \left[\sum_{a=1}^{N}\psi_{a}^{\dagger}\left(\frac{d}{dt} - \mu\right)\psi_{a} - \sum_{a,b,c,d=1}^{N}J_{ab;cd}\,\psi^{\dagger}_{a}\psi^{\dagger}_{b}\psi_{c}\psi_{d}\right]\,.
   \label{csyk}
\end{eqnarray}
Here, $\mu$ is a chemical potential that tunes the $U(1)$ charge $\displaystyle Q = \sum_{a=1}^{N}\langle \psi^{\dagger}_{a}\psi_{a}\rangle$ and the $J_{ab;cd}$ are now complex, independent, random couplings drawn from a Gaussian distribution with zero mean and that satisfy
\begin{eqnarray*}
   J_{ab;cd} &=& - J_{ba;cd} = -J_{ab;dc} = J^{*}_{cd;ab}\,,\\
   \overline{|J_{ab;cd}|^{2}} &=& J^{2}\,.
\end{eqnarray*}
In addition, the fermions close the algebra $\{\psi_{a},\psi_{b}\} = 0$, $\{\psi_{a},\psi_{b}^{\dagger}\} 
=\delta_{ab}$.  This complex SYK model shares many properties with its more well-understood real counterpart and that we will discuss in more detail in section 3 below. Our original goal in this article was to gauge the global $U(1)$ by coupling the complex fermions to a $U(1)$ gauge field $A(t)$ and investigating its properties. In particular, we set out with the aim of understanding if the resulting model,
\begin{itemize}
   \item possesses a sensible classical gravitational dual,
   \item retains the chaos structure of the original (complex) SYK model and,
   \item is sensitive to the breaking of the reparameterization invariance as the theory is moved away from its conformal limit.
\end{itemize}
Importantly, in (0+1)-dimensions the gauge field is non-dynamical since a Maxwell kinetic term cannot be added to the action but, being an odd-dimensional spacetime means that a Chern-Simons term can. In this case it is simply linear in $A(t)$. The result is a form of topological quantum mechanics \cite{Dunne:1989hv, Floreanini:1991iu, Jackiw:1997kp} originally studied by Dunne et.al. in the context of finite temperature Chern-Simons theory \cite{Dunne:1996yb}, with an added layer of random interactions.  However\footnote{We would like to thank the anonymous referee for highlighting this point to us.}, as we will show below, this coupling to an {\it external} gauge field - {\it i.e.} one that is not integrated over in the path integral - is equivalent to a twisting of the fermion boundary conditions on the thermal circle.\\

\noindent
This twisting of the boundary conditions of random mass, and/or randomly interacting fermions, is expected to introduce novel effects to the disorder and interactions that can drastically change the behavior of the system. Depending on the parameters of the model, such as the mass distribution, the ``Chern-Simons'' coupling, and the chemical potential, the system can exhibit different phases with different properties, such as localization, delocalization, Luttinger liquid behavior, and gapping \cite{ Chowdhury:2021qpy}. These phases can be characterized by various physical observables, such as the energy levels, the charge density, the gap equation, and the density of states. As a warmup to the problem we really want to study, {\it viz} the charged SYK$_{4}$ model with twisted boundary conditions in one spacetime dimension, in this article we turn our attention to the simpler, integrable, quadratic SYK model. Even in this much simplified setting, we show that there are some interesting lessons to be learnt, lessons that we will port to the canonical $q=4$ model in forthcoming work.\\

\noindent
The SYK$_2$ model, with a disordered 2-fermi interaction term, is equivalent to a theory of free fermions with a random mass matrix. As we shall see, using the standard SYK approach to formulate the theory in terms of the collective field variables, the Schwinger-Dyson equations may be combined into a single quadratic equation. This may then be solved for the large $N$ two-point correlation function for arbitrary coupling strength. This is in contrast to the $q>2$ cases of the model, in which one can only find an analytic expression for the large $N$ two-point function in the strong-coupling/low-energy conformal limit. This great simplification makes the SYK$_2$ model particularly amenable to analytic study, where it has been shown to have the following basic features: firstly, it is an integrable model, as the OTOC does not exhibit the exponential growth characteristic of quantum chaos \cite{Gross:2016kjj}. Secondly, it has an average spectral density given by the Wigner semicircle law in the large $N$ limit \cite{Gross:2016kjj,Cotler:2017jue}, just like the Gaussian RMT ensembles. However, despite these seemingly trivial features, the model has proven to be a remarkably rich object of study.
\\

\noindent
The rest of the article is structured as follows: In the interests of pedagogy, in Section 2 we introduce the coupling of the complex SYK model to the external ``Chern-Simons'' gauge field {\it a la} \cite{Dunne:1996yb,Dunne:1989hv}, and show how the resulting model is equivalent to twisting the fermion boundary conditions on the thermal circle. Section 3 is taken up with describing some of the technicalities of the disorder averaging procedures we use in the rest of the paper. The crux of the paper is contained in Sections 4 and 5 where we compute the out-of-time-order correlator (or, more precisely, the double commutator) and the spectral form factor of the model respectively. We summarise our results and comment on future directions in Section 6. Finally, we collect some technical details of our computations such as the derivation of the 2-replica spectral form factor, the partition functions and the mass matrices in the appendices.

\section{Chern-Simons quantum mechanics}

Our starting point is the following Lagrangian \cite{Dunne:1989hv, Floreanini:1991iu, Dunne:1996yb, Jackiw:1997kp}, written in the imaginary time formalism, is given by
\begin{equation}
L = \sum_{j=1}^{N_f} \psi_j^\dag ( \partial_\tau - i A + m ) \psi_j \,,  \label{LDunne}
\end{equation}
where the fermions close the algebra
\begin{equation}
\left\{ \psi_i^\dag, \psi_j  \right\} = \delta_{ij}.  
\end{equation}
$U(1)$ gauge transformations act on the fields as
\begin{eqnarray}
\psi & \rightarrow & e^{i \omega(\tau)}    \psi \nonumber \\
A & \rightarrow & A + \partial_\tau \omega.
\end{eqnarray}
Note that the path integral corresponding to \eqref{LDunne} is invariant under the gauge transformation.  In order to maintain the antiperiodicity of the gauge transformed fermion fields, the phase must satisfy
\begin{align}\label{gaugeinv}
    \omega(\beta)-\omega(0) = 2\pi k, \quad k\in\mathbb{Z}.
\end{align}
Transformations for which $k\neq0$ are called large gauge transformations. This gauge freedom means that we need only consider constant gauge field configurations since, for any $A(\tau)$ we may always make the gauge transformation
\footnote{Note that this is a small gauge transformation since $\omega(\beta) = \omega(0).$ }\begin{align}\label{gaugetrans}
    \omega(\tau) = \frac{a}{\beta}\tau - \int_0^\tau d\tau' A(\tau'),
\end{align}
where $a = \int_0^{\beta} d\tau A(\tau)$, to get to the constant configuration
\begin{align}\label{constgauge}
    A(\tau) \rightarrow \frac{a}{\beta}.
\end{align}
It should be emphasised that, although the gauge field coupling naively resembles a constant chemical potential, its transformation properties are different.  In particular, all expressions in the imaginary time or real-time formalism should be invariant under the large gauge transformations of the form $a \rightarrow a + 2 \pi k $ for integer $k$.
\\ \\
Furthermore we emphasize that, since the gauge field is not being integrated in the path integral, one could redefine the fermion fields as 
\begin{equation}
\psi'(\tau) = \psi(\tau) e^{-i \int_0^\tau d\tau' A(\tau') }    \nonumber
\end{equation}
which eliminates the gauge field in (\ref{LDunne}).  Where the fields $\psi$ satisfy the usual antiperiodicity condition, the fields $\psi'$ satisfy
\begin{equation}
\psi'(\beta) = e^{-i a} \psi'(0).    \nonumber
\end{equation}
As such, the introduction of the gauge field is equivalent to introducing a twisted boundary condition. 
\\ \\
One of the many features of this model is that its Green's function  can be computed analytically. In particular in the $m\rightarrow 0$ limit, 
\begin{equation}
\langle \psi_i^\dag (\tau_1) \psi_j(\tau_2) \rangle \equiv G^0_{ij}(\tau_1 - \tau_2) = \left\{  \begin{array}{cc}  \frac{\delta_{ij}}{2}\left( -1 - i \tan\left( \frac{ a}{2}  \right)  \right) e^{i \frac{ a}{\beta} (\tau_1 - \tau_2)}     & -\beta < \tau_1 - \tau_2 < 0    \\ 
- \frac{i \delta_{ij}}{2} \tan\left( \frac{ a}{2}  \right)   &   \tau_1 - \tau_2 = 0 \\ 
\frac{\delta_{ij}}{2}\left( 1 - i \tan\left(\frac{ a}{2} \right)  \right) e^{i \frac{ a}{\beta}(\tau_1 - \tau_2)}   &  0 < \tau_1 - \tau_2 < \beta   \\
 \frac{i \delta_{ij}}{2} \tan\left(  \frac{ a}{2}  \right)   &   (\tau_1 - \tau_2) = \beta.
\end{array}    \right.   \label{G0Propagator}
\end{equation}
As explained in some detail in \cite{Dunne:1996yb}, the Lagrangian  (\ref{LDunne}) can be derived as a reduction of (2+1)-dimensional Chern-Simons matter \cite{Jackiw:1997kp}, and provides a simplified toy model in which some of the subtleties of its higher-dimensional avatar may be clarified. 
 Inspired by the Sachdev-Ye-Kitaev model \cite{Sachdev:1992fk}, we will instead be interested in perturbing the model with random $q$-body fermion operators. In this sense, this is an SYK$_{q}$ model coupled to an external gauge field.   As noted in the introduction, the SYK$_q$ model has a number of interesting features such as its exact solvability at large $N$, and the fact that it is maximally (quantum) chaotic \cite{Maldacena:2016hyu, Polchinski:2016xgd, Garcia-Garcia:2016mno, Gross:2016kjj,Gross:2017aos, Saad:2018bqo,  Rosenhaus:2018dtp}.  The latter suggests that it should admit a holographic description as a black hole (in two dimensions) \cite{Penington:2019kki, Trunin:2020vwy}, while the former allows (in principle) for details of the holographic dictionary to be systematically unpacked.\\ \\
A special case of this system is when $q=2$ where the model is non-chaotic and may be solved exactly \cite{Gross:2016kjj} and so serves as an important test case for the gauged SYK model. In particular, we would like to understand whether the inclusion of the gauge field spoils its solvability properties and/or introduces chaotic regions in its parameter space\footnote{In particular, it would be interesting to explore whether integrable-chaotic transitions are present in parameter space \cite{Sorokhaibam:2019qho}.}. To be explicit, we will focus on the model,
\begin{equation}
L = \sum_{j=1}^{N_f} \psi_j^\dag ( \partial_\tau - i A ) \psi_j  +   \sum_{ij} \psi^\dag_i m_{ij}  \psi_j   \label{CSQMwithMass}
\end{equation}
where the random mass matrix $m_{ij} = m_{ji}^{*}$ is drawn from a Gaussian distribution with zero mean, 
\begin{equation}\label{Pmij}
P(m_{ij}) = \left\{ \begin{array}{cc}  (\frac{N}{\pi m^2} )^{\frac{1}{4}}  e^{- \frac{N}{2m^2} |m_{ij}|^2 }  &   i \neq j  \\
 \sqrt{\frac{N}{ 2 \pi m^2} } e^{- \frac{N}{ 2m^2} m_{ii}^2}  &   i = j  \end{array} \right.
\end{equation}
and variance,
\begin{equation}
\overline{m_{ij} m_{ji}} = \overline{ m_{ii}^2 } = \frac{m^2}{N}\,.
\end{equation}
Here we have defined $\displaystyle \overline{F} = \int F \prod_{i \neq j } P(m_{ij}) dm_{ij} \prod_i P(m_{ii}) dm_{ii}$, and use an overline to denote the disorder averaging throughout. 
 In addition, we need to specify whether we consider an annealed or quenched averaging. In the former we perform the disorder averaging at the beginning, obtaining a new post-averaging action while in the latter we compute correlation functions for mixed choices of $m_{ij}$ and perform the averaging afterwards.  In this article we will utilise both approaches.  Our computation of the spectral form factor will use annealed averaging while the out-of-time-ordered correlator will use quenched averaging. At large $N$, we expect these results to coincide \cite{Baldwin2020}, but hope that the development of both approaches will prove useful when different perturbations of (\ref{LDunne}) are considered.  


\section{Disorder averaging}

\label{QuenchedAverageSection}

Before presenting our findings for the SFF and OTOC for the model described by (\ref{CSQMwithMass}), it will be useful to take note of some results that are relevant for the computation of general correlation functions of disordered system. 

\subsection{Quenched averaging}

\begin{figure}
\centering
\includegraphics[width=0.90\textwidth]{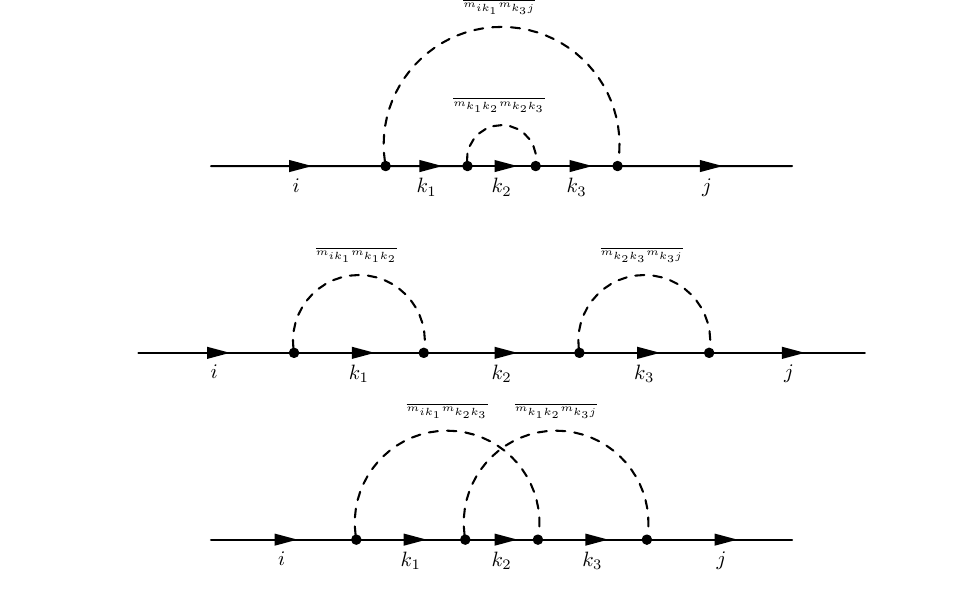}
\caption{The diagrams relevant for four mass insertions.  The fermion propagators are indicated by solid lines and are directed from $\psi^\dag$ to $\psi$.  The dashed lines are related to the mass-averaging, which require the masses to be paired up to give a non-zero result.  The top two diagrams are rainbow diagrams for which the dashed lines do not intersect.  Diagrams such as this represent the leading large $N$ contribution since the smallest number of flavor indices are contracted.}
\label{fourMasses}
\end{figure}

\begin{figure}
\centering
\includegraphics[width=0.90\textwidth]{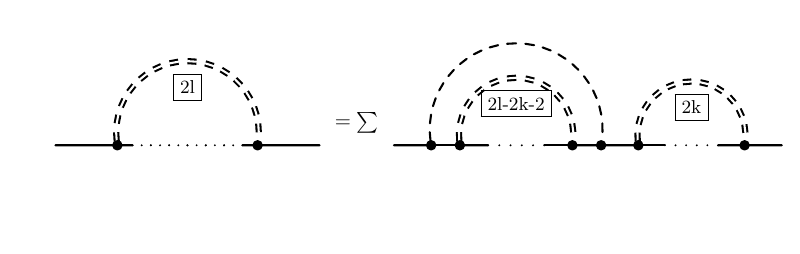}
\vspace{-1.0cm}
\caption{The number of rainbow diagrams for $2l$ mass insertions can be determined by a recursive relation, displayed here in diagrammatic form.  The double dashed lines sum over all rainbow diagrams involving the number of insertions in the square box. }
\label{Rainbows2}
\end{figure}

We begin by collecting results that will be useful for carrying out a quenched averaging prescription. In this case, all Feynman diagrams consist of integrals of the two-point function (\ref{G0Propagator}) with some collection of $m_{ij}$ terms inserted as vertices. As such, the convolution of $n$ two-point functions is given by
\begin{eqnarray}
   & &  \int_0^{\beta} d\tau_1 \int_0^\beta d\tau_2 \cdots \int_0^\beta d\tau_n G^0_{k_1 k_1}(t_1 - \tau_1) G^0_{k_2 k_2}(\tau_1 - \tau_2) \cdots G^0_{k_n k_n}(\tau_n - t_2)   \nonumber \\
   &=& e^{i \frac{ a}{\beta}(t_1 - t_2) } \sum_{p=0}^n  \frac{(t_1 - t_2)^{n-p} \beta^p (-1)^p  }{2^{p+1} p! (n-p)!} \left. \partial^{\,p}_{x} \left( \tanh x  + \mathrm{sgn}(t_1 - t_2) \right)   \right|_{x \rightarrow \tanh^{-1}(-i \tan(\frac{ a}{2}  ))}    \nonumber
\end{eqnarray}
where $G^{0}_{ij}$ is defined in (\ref{G0Propagator}).  The mass averaging can be carried out for arbitrary collections of mass insertions.  We can do the integrals separately for the diagonal couplings, and pairwise separately for the off-diagonal ones.  Consequently,
\begin{eqnarray}
\overline{m_{ij}^s m_{ji}^r } & = &  r \frac{m^{2r}}{N^r}  \Gamma(r) \delta_{s,r} \ \ \ i \neq j \nonumber \\
\overline{m_{ii}^{2r}} & = &  2^r \frac{m^{2r}}{N^r} \frac{\Gamma(r + \frac{1}{2})}{\sqrt{\pi}}  \ \ \ r \in N_0 \\
\overline{m_{ii}^{2r+1}} & = &  0 \ \ \ r \in N_0 \nonumber
\end{eqnarray}
Since the two-point function $G^0_{ij}$ is proportional to $\delta_{ij}$, this forces the indices of the mass insertions to order into a sequence. Consequently, for all diagrams one has to perform averages of the form
\begin{equation}
  \sum_{k_i =1}^N \overline{ m_{i k_1} m_{k_1 k_2} \cdots m_{k_{s} j}   }
\end{equation}
 In order for this quantity to be non-zero, the number of $m_{ij}$ terms must match the number of $m_{ji}$ terms for all $i,j$.  The leading large $N$ contribution comes from the so-called ``rainbow diagrams'' \cite{Gross:2016kjj}.  These are illustrated in Fig. (\ref{fourMasses})
and result in
\begin{equation}
  \sum_{k_i =1}^N \overline{ m_{i k_1} m_{k_1 k_2} \cdots m_{k_{2l-1} j}   } = \frac{1}{l+1}\left( \begin{array}{c} 2l \\ l  \end{array}  \right) m^{2l} \delta_{ij} + \mathcal{O}( \frac{1}{N} )   \label{largeNmass}
\end{equation}
The combinatoric factor can be computed by solving a recursive equation, see Fig. (\ref{Rainbows2}).  

\subsection{Imaginary time two-point function}

Remarkably, these results may be combined to give a closed form expression for the fully dressed two-point correlation function.   The diagrams we need to sum correspond to 
\begin{eqnarray}
& & \overline{\langle \psi_{i}^\dag(\tau) \psi_{j} \rangle }  \equiv  \bar{G}_{i j}(\tau) \nonumber \\ & = & \sum_{l=0}^{\infty} \int_0^{\beta} d\tau_1  \cdots \int_0^\beta d\tau_{2l} \ G^0_{i i}(\tau - \tau_1) G^0_{k_1 k_1}(\tau_1 - \tau_2) \cdots G^0_{k_{2l-1} k_{2l-1}}(\tau_{2l-1} - \tau_{2l}) 
 G^0_{jj} (\tau_{2l})  \times 
\nonumber \\
&  &  \left(     \sum_{k_i =1}^N \overline{ m_{i k_1} m_{k_1 k_2} \cdots m_{k_{2l-1} j}   } \right)\,. 
\end{eqnarray}
By taking the large $N$ limit and making use of an integral representation of the hypergeometric function we obtain 
\begin{eqnarray}
\overline{G}_{i_1 j_2}(\tau) 
&=& \delta_{j_2 i_1} e^{\frac{i a}{\beta}\tau} \frac{1}{\pi} \int_{-1}^1 dy \sqrt{1 - y^2} e^{-2 m y \tau} \left( \tanh\left(-i \frac{ a}{2}  + m y \beta\right) + \textnormal{sgn}(\tau)  \right)  \nonumber \\
& = &  \delta_{j_2 i_1} \textnormal{sgn}(\tau) e^{i \frac{ a}{\beta} \tau} \frac{2}{\pi} \int_{-1}^{1} dy \sqrt{1 - y^2} \frac{e^{-2 m y \tau}}{ 1 +  e^{-2 \textnormal{sgn}(\tau) m y \beta    } e^{ i\textnormal{sgn}(\tau) a   }   }   \nonumber \\
&=& \delta_{j_2 i_1} \textnormal{sgn}(\tau)  \frac{2}{\pi} \int_{-1}^{1} dy \sqrt{1 - y^2} \frac{e^{-\left(2 m y  + 
  i \frac{ a}{\beta} 
 \right)\tau}}{ 1 +  e^{- \textnormal{sgn}(\tau) \left(2 m y  + 
  i \frac{ a}{\beta} 
 \right)\beta   }   } \,.   \label{GFullyDressed}
\end{eqnarray}
Taking $0 <\tau < \beta$ and transforming the $y$-coordinate, we obtain
\begin{eqnarray}
\overline{G}_{i_1 j_2}(\tau) & = &  \delta_{j_2 i_1} e^{i \frac{ a}{\beta} \tau} \int_{-2m}^{2m} dy \frac{\sqrt{4 m^2 - y^2}}{2 \pi m^2} \frac{e^{-y \tau}}{ 1 +  e^{- y \beta    } e^{ i a   }   }\,,     \label{GmSummed}
\end{eqnarray}
which clearly reproduces the correct limit for $ a \rightarrow 0$, (see Appendix C of \cite{Gross:2016kjj}) and is invariant under large gauge transformation\footnote{Note that the overall phase accounts for the transformation of the fermion fields under the gauge transformation}.  
On the other hand, when $m\beta >> 0$, the integral may be computed analytically.  The leading contribution to this integral is given by
\begin{eqnarray}\label{leadingpiece}
\overline{G}_{i_1 j_2}(\tau) & = &  \delta_{j_2 i_1} \textnormal{sgn}(\tau) e^{i \frac{ a}{\beta} \tau} \frac{2}{\pi} \int_{0}^{1} dy \sqrt{1 - y^2} e^{-2 m y | \tau |}     \nonumber \\
 &=& \frac{\delta_{j_2 i_1} }{2 m \tau} e^{i \frac{ a}{\beta} \tau} \left( I_1(2 m |\tau|) - L_{1}(2 m |\tau |)    \right)\,,
\end{eqnarray}
where $I_1$ is the modified Bessel function and $L_1$ is the modified Struve function and differs from the expression C.7 in \cite{Gross:2016kjj} only by an overall phase.   For general $m\beta$ we can write 
(\ref{GmSummed}) as
\begin{equation}
\bar{G}_{i_1 j_2}(\tau) = \delta_{j_2 i_1} \frac{2}{\pi \beta} \int_{-1}^1 dy \sqrt{1 - y^2} \sum_{n=-\infty}^{\infty} \frac{e^{i (2n-1) \frac{\pi}{\beta} \tau  }}{\frac{(2n-1)\pi i}{\beta} + (2 m y  + 
  i \frac{ a}{\beta} 
 )}
\end{equation}
after making use of an identity in \cite{Dunne:1996yb}.  This expression can also be obtained by performing the inverse discrete Fourier transform and setting $\omega_n = \frac{\pi}{\beta}(2 n -1)$ to give,
\begin{eqnarray}
\bar{G}_{j_2 i_1}(\omega_n) = \frac{2}{\pi} \delta_{j_2, i_1} \int_{-1}^1 dy \sqrt{1 - y^2} \frac{-1}{i \omega_n - (2 m y  + 
  i \frac{ a}{\beta} 
 ) }   \nonumber \\
\end{eqnarray}
Note that while the above expression for an individual mode is not invariant under the large gauge transformation $a \rightarrow a+ 2 \pi k$, the set of {\it all} modes is.

\subsection{Real-time two-point function}

Given $\overline{G}_{j_{2}j_{1}}(\omega_{n})$, the real-time retarded Green's function is usually obtained by analytically continuing $i \omega_n \rightarrow \omega + i \epsilon$.  In addition, however, we need to Wick rotate the gauge field $a \rightarrow -i a$ as we are transforming from real to imaginary time. To disambiguate between the real- and imaginary time expressions we also replace $\beta \rightarrow T $.  This step is essential for preserving the gauge invariance of our expressions.  Then, taking the Fourier transform and carrying out the integral over $y$, we obtain
\begin{eqnarray}
\bar{G}_{j_{2} i_1}(t) & = & \frac{2}{\pi} \delta_{j_2, i_1} \int \frac{d \omega}{2\pi }  \int_{-1}^1 dy \sqrt{1 - y^2} \frac{-1}{\omega + i \epsilon - (2 m y  + 
   \frac{ a}{T} 
 ) } e^{-i \omega t}    \nonumber \\ 
 &=& i e^{-i \frac{ a}{T} t} \frac{J_{1}(2 m t)}{m t} \theta( t)\,,   \label{RealTimeG}
 \end{eqnarray}
 where $J_{1}$ is a Bessel function of the first kind.  This agrees with the corresponding expression in \cite{
Michel:2016kwn} up to an additional phase factor in $t$, which does not affect the late-time decay of the expression $\frac{J_1 ( 2 m t )}{m t} \sim t^{-\frac{3}{2}}$. Without simultaneously Wick rotating the gauge field, we would instead obtain an overall (decaying) exponential factor which breaks the gauge invariance.   The answer (\ref{RealTimeG}) may also be obtained by deforming the integration contour \cite{
Michel:2016kwn}. By first performing the integral over $y$ followed by the analytic continuation and Wick rotation we obtain
\begin{equation}
\bar{G}_{j_2 i_1}(\omega) =  -\delta_{j_2 i_1} \left(\omega - \frac{ a}{T}\right)\frac{1 -\sqrt{1 - \frac{4 m^2}{ (\omega -  \frac{ a}{T}  + i \epsilon)^2}}  }{ 2 m^2 }   \label{realTimeFT}
\end{equation}
The square root factor introduces a branch cut that is shifted just off the real axis.  For positive time we close the contour in the lower complex plane and get a non-vanishing result since the branch cut is contained in the contour.  The contour may be deformed to one that  hugs the branch cut, yielding  (\ref{RealTimeG}). The combination $\frac{J_1 ( 2 m t )}{m t}$ decays as $t^{-\frac{3}{2}}$ at late times. 

\subsection{Annealed averaging}

The choice of a Gaussian distribution for the random coupling makes it particularly convenient to carry out an annealed averaging. Depending on the quantity of interest, a number of replicas may be constructed, the random couplings for the replicas identified and the Gaussian integral may be performed exactly.  The general integrals we will be interested in are simple Gaussians, 
\begin{eqnarray}
&&\int \prod_{i \neq j } P(m_{ij})\, dm_{ij} \prod_k P(m_{kk})\, dm_{ii}
 \exp\Biggl[ m_{ij} \left( \sum_\alpha w_\alpha \int d\tau 
 \psi_{i_\alpha}^\dag \psi_{j_\alpha} \right)\nonumber\\  
 &+&  m_{kk} \left(\sum_\alpha w_\alpha \int d\tau 
 \psi_{k_\alpha}^\dag \psi_{k_\alpha}  \right)\Biggr]\,,  
\end{eqnarray}
where the label $\alpha$ runs over  replicas.  The weight factor $w_\alpha$ is relevant when computing real-time correlators and complex conjugates.  \\ \\
In Appendix \ref{appendix:A} we compute the disorder average for the $2$-replica spectral form factor, while in Appendix \ref{appendix:B} we compute the partition function, involving a single replica.  At large $N$ we find the expected agreement between the two-point function computed using either the annealed or quenched prescriptions.   

\section{Double commutator}

The out-of-time-ordered correlator (or the closely-related double commutator correlation function) is a widely used contemporary diagnostic of quantum chaos \cite{Larkin1969QuasiclassicalMI, Almheiri:2013hfa, Shenker:2013pqa, Shenker:2013yza, Roberts:2014isa, Roberts:2014ifa, Stanford:2015owe, Maldacena:2015waa}.  It will serve a similar purpose in our study of the model (\ref{CSQMwithMass}). Specifically, the correlator
\begin{equation}
    C(t) = \theta(t) \langle \sqrt{\rho} \left\{ \psi^\dag(t)_i, \psi(0)_j \right\}^\dag  \sqrt{\rho}  \left\{ \psi^\dag(t)_i, \psi(0)_j \right\}   \rangle   \label{doubleCommutator}
\end{equation}
is expected to exhibit exponential growth at early times for chaotic systems \cite{Maldacena:2015waa}.  The rate of exponential growth is characterised by the spectrum of Lyaponov exponents.  The splitting of the factor $\rho$ into the two factors of $\sqrt{\rho}$ regulates some divergences and places the insertions on opposite sides of the thermal circle \cite{Stanford:2015owe}.  \\ \\
To compute $C(t)$ (in the quenched averaging prescription) we need to perform the standard summing over all Feynman diagrams.  As emphasised in section \ref{QuenchedAverageSection}, this involves the sum of convolutions of the undressed two-point function, which follows from the Wick contractions.  The structure of the Wick contractions for $C(t)$ has a special feature due to the anti-commutators.  Specifically, as a result of the fermionic statistics, contractions involving an anti-commutator of fields vanish unless the fields in the anti-commutator are contracted with each other. At tree level this implies factorisation of (\ref{doubleCommutator}) into two retarded propagators,  
  a feature which persists even beyond tree level where perturbative two-body insertions may always be combined into an anti-commutators of fields. This implies that, for our model, Wick contractions can only occur between fields that are inserted on the same side of the thermal circle, dramatically simplifying (\ref{doubleCommutator}) to
\begin{eqnarray}
    C(t) & = & \theta(t) G_{ij}(t) (G_{ij}(t))^*    \nonumber \\
    G_{ij}(t_1 - t_2) &=& \sum_{n = 0 }^\infty \int dt_1' dt_2' \cdots dt_n' G^0_{i i}(t_1 - t_1') G^0_{k_1 k_1}(t_1' - t_2') \cdots G^0_{j j}(t_n' - t_2) \times    \nonumber \\
    & & \sum_{k_i=1}^N m_{i k_1} m_{k_1 k_2} \cdots m_{k_{n-1} j}\,,    
    \label{CtFactorise}
\end{eqnarray}
where we have used the fact that $G^0_{ij} \propto \delta_{ij}$.  We did not, however, impose any further conditions on $G^0_{ij}$ nor did we need to use the disorder averaging of the mass terms.  The result (\ref{CtFactorise}) only makes use of the fact that we have perturbative two-body interactions and the special form of the Wick contractions for anti-commutators of fields.   \\ \\
The results of section \ref{QuenchedAverageSection} can be used directly to simplify the above expression.  However, a comment of caution is in order.  At first glance the above expression looks like the absolute-value-squared of the dressed two-point function however, unlike the Wick contractions, the mass averaging need not involve mass insertions from the same side of the thermal circle.
\\ \\
To leading order in $1/N$ then, we can write
\begin{equation}
   \bar{C}(t)  =  \theta(t) \bar{G}_{ij}(t) (\bar{G}_{ij}(t))^*   +    \frac{\theta(t)}{N}\bar{F}_{ijji}(t) + O(\frac{1}{N^2})
\end{equation}
so that at leading order in $N$, the double commutator factorises into the product of dressed two-point functions.  The $1/N$ corrections come in two forms.  The first are of the kind in (\ref{largeNmass}) which will further dress the two-point function. The second come from corrections involving mass averaging on opposite sides of the thermal circle.  We will focus on the latter since these give rise to ladder diagrams, the proliferation of which is associated with the spectrum of growth exponents \cite{Stanford:2015owe}.    \\ \\
We may regroup the terms in (\ref{CtFactorise}) to  isolate the masses that are involved in averages on opposite sides of the thermal circle.  The sum over mass insertions for masses averaged on the same side of the thermal give rise to dressed two-point functions. This means that
\begin{eqnarray}
\frac{1}{N}F_{i j j i}(t) & = & \sum_{r=1}^\infty \left( \int dt_1 dt_2 \cdots dt_n \bar{G}_{i i}(t -t_1) \bar{G}_{k_1 k_1}(t_1 - t_2) \cdots \bar{G}_{j j}(t_{r-1})  \right)  \times   \nonumber \\
& & \left( \int dt'_1 dt'_2 \cdots dt'_n \bar{G}_{i i}(t -t'_1) \bar{G}_{l_1 l_1}(t'_1 - t'_2) \cdots \bar{G}_{j j}(t'_{r-1})  \right)^*  \times \nonumber \\
& & \sum_{k_i, l_i = 1}^N (m_{i k_1} m_{k_1 k_2} \cdots m_{ k_{r-1} j} )( m_{i l_1} m_{l_1 l_2} \cdots m_{ l_{r-1} j }  )^* \,,
\end{eqnarray}
where we have isolated the $r$ masses involved in averaging on opposite sides of the thermal circle. Taking the Fourier transform and isolating the leading large $N$ contribution, we find
\begin{eqnarray}
\bar{F}_{i j j i}(t) & = & 2 \delta_{i j} \sum_{r=1}^\infty \int d\omega_1 d\omega_2 e^{-i (\omega_1 - \omega_2) t} \left( 
 \bar{G}(\omega_1) (\bar{G}(\omega_2) )^*  \right)^{r+1} m^{2r}    \nonumber \\
 & = & 2 \delta_{i j} \int d\omega_1 d\omega_2 e^{-i (\omega_1 - \omega_2) t}  \frac{m^2 \left( 
 \bar{G}(\omega_1) (\bar{G}(\omega_2) )^*   \right)^2 }{1 - m^2 \bar{G}(\omega_1) (\bar{G}(\omega_2) )^* }     \nonumber \\
 &=& 2 \delta_{ij} \int d\omega_1 d\omega_2 e^{-i (\omega_1 - \omega_2) t} \left(  
 -\bar{G}(\omega_1) (\bar{G}(\omega_2) )^*    +  \frac{1  }{1 - m^2 \bar{G}(\omega_1) (\bar{G}(\omega_2) )^* }      \right)\,,   \nonumber
\end{eqnarray}
in agreement with the results of \cite{Michel:2016kwn, Gross:2016kjj}.  The Green's function in the above expression is the Fourier transform of the real-time Green's function (\ref{realTimeFT}). 
\\ \\
By shifting the integration variables $\omega_j \rightarrow \omega_j + \frac{ a}{T}$, we obtain an integral of the type considered in \cite{Gross:2016kjj}. The phase factors from the two-point functions cancel out and thus the gauge field has no effect on the double commutator. 
As pointed out in \cite{Polchinski:2016xgd, Gross:2016kjj}, the resulting expression does not give rise to exponential growth of the OTOC so that we conclude that our model is not chaotic. This is not surprising since neither is the underlying SYK$_{2}$ model.

\section{Spectral form factor}
A hallmark of chaotic systems is that they exhibit spectral rigidity, in the sense that the eigenvalues of the Hamiltonian repel each other {\it i.e.} they forget as much information as possible, modulo what they are compelled to remember by their symmetries. This heuristic picture  is succinctly encoded in the so-called Bohigas-Giannoni-Schmit (BGS) conjecture \cite{Bohigas:1983er} which holds that any quantum chaotic system should exhibit the spectral statistics of a random matrix ensemble consistent with the time-reversal symmetries of the model. One very effective diagnostic tool for this random matrix theory (RMT) behaviour is the {\it spectral form factor} (SFF), the Fourier transform of the 2-point correlator of eigenvalues of the Hamiltonian. At the level of the SFF, a signature of the level-repulsion indicating RMT behaviour, is the emergence of a linear scaling with time - the so-called linear ramp - at intermediate times \cite{Cotler:2016fpe, Saad:2018bqo, Liu:2018hlr, Cotler:2017jue}.\\ 

\noindent
 The SFF also encodes specific non-chaotic features of disorder-averaged integrable many-body systems, such as the (usual Majorana) SYK$_2$ model considered in \cite{Winer:2020mdc}. There, the authors show that the 2-replica action possesses an unusual, large $SU(2)$ symmetry, which gives rise to an exponential, as opposed to a linear, ramp in the SFF. They further conjecture that this exponential ramp should be a feature of {\it all} non-interacting disordered systems, and attribute this behaviour of the SFF to the presence of a high-dimensional manifold of saddle points resulting from the symmetry group. The gauged SYK$_2$ model which we consider in this article provides another setting in which to test these claims.\\

 \noindent
 We note in passing that the complex SYK$_2$ model with chemical potential has been considered previously in \cite{Liao:2020lac}, largely using  different methods to the path integral approach utilised here as well as in \cite{Winer:2020mdc}. In real time - in which the SFF is calculated - our model is superficially equivalent to theirs with the important distinction that the gauge field transforms differently to a chemical potential and acquires a temperature dependence in constant gauge (\ref{constgauge}).\\
 
 \noindent 
 Practically speaking, the spectral form factor can be computed by analytically continuing the partition function $Z=Z(\beta)$ to $Z(\beta+iT)$. The SFF is subsequently defined as
\begin{equation}
    \label{SFFdef}
    g(\beta;T) \equiv \frac{\overline{Z(\beta+iT)Z^*(\beta+iT)}}{\overline{Z(\beta)}^{2}}\,.
\end{equation}
Its disconnected and connected components are defined respectively by
\begin{eqnarray}
    g_d(\beta;T) &=& \frac{\overline{Z(\beta+iT)} \ \overline{Z^*(\beta+iT)}}{\overline{Z(\beta)}^2}, \label{gd}\\
    g_c(\beta;T) &=& g(\beta;T) -g_d(\beta;T)\,.
\end{eqnarray}
Note that according to these definitions we are treating the disorder as annealed, and not quenched. This allows us to perform the disorder average over the denominator and numerator separately.  \\ \\
We will work primarily in the infinite temperature limit $g(\beta=0;T)\equiv g(T)$. To calculate $Z(iT)$, we simply Wick rotate from Euclidean to real time, under which $\tau\rightarrow it$, $\ A(\tau)\rightarrow -iA(t)$ and integration runs from 0 to $T$. Working in the constant gauge $A(\tau)=a/\beta$ which in real time becomes $A(t)=a/T$, our action,
\begin{align}\label{SFFaction}
\tilde{I} = iS = -\int_0^T dt \left[ \sum_{i=1}^N \psi_i^\dag \left(\frac{d}{dt} - i\frac{a}{T} \right) \psi_i +i \sum_{i,j=1}^N m_{ij} \psi_i^\dag \psi_j \right].
\end{align}
As is usual in the path integral  computation of the SFF, we consider the physics of fermions in \textit{periodic} real time, with period $T$ and with the fermions satisfying anti-periodic boundary conditions.  \\

\noindent
Averaging over the disorder (see Appendix \ref{appendix:A} for more details), yields the following exact expression in terms of bilocal, collective fields $G$ and $\Sigma$,
\begin{eqnarray}
g(T) &\equiv& \overline{Z(iT) Z(iT)^*}=\prod_{\alpha,\beta} \int \mathscr{D}\Sigma_{\alpha \beta} \mathscr{D}G_{\alpha \beta} e^{N I(G,\Sigma)}\,, \label{SFFeff1}
\end{eqnarray}
where,
\begin{eqnarray}
I(G,\Sigma) &=& \log\det\left[ - \delta(t-t')\left(\delta_{\alpha \beta}\frac{d}{dt} - i \frac{a}{T}\sigma^z_{\alpha\beta} \right) + \Sigma_{\beta\alpha}(t,t') \right] \label{Icomplexa}
\\
&& - \int dt dt' \sum_{\alpha,\beta} \left[ \Sigma_{\beta\alpha}(t',t) G_{\alpha\beta}(t,t') -(-1)^{\alpha+\beta} \frac{m^2}{2} G_{\alpha \beta}(t,t')G_{\beta\alpha}(t',t) \right], \nonumber
\end{eqnarray}
and $\sigma^z$ is the usual Pauli $z$-matrix. Greek indices run over the two replicas. In order to tackle this path integral we will assume that time-translation invariant configurations of $G$ and $\Sigma$ dominate \cite{Winer:2020mdc}. 
 This allows us to take a discrete Fourier transform
\begin{align}
    G_{\alpha\beta}(t) &= \frac{1}{T}\sum_{n \ \text{odd}} G_{\alpha\beta}(\omega_n) e^{-i\omega_n t}, \quad
    \Sigma_{\alpha\beta}(t) = \frac{1}{T}\sum_{n \ \text{odd}} \Sigma_{\alpha\beta}(\omega_n) e^{-i\omega_n t}, \qquad \omega_n=\frac{n\pi}{T},
\end{align}
and ensures that variables with different Matsubara frequencies decouple.  In terms of these variables the path integral becomes 
\begin{equation}
g(T) = \prod_{n \ \textnormal{odd}} \prod_{\alpha, \beta}  \left( \int  d\Sigma_{\alpha \beta}(\omega_n) dG_{\alpha \beta}(\omega_n)  e^{N I_n(G(\omega_n), \Sigma(\omega_n))}  \right)
\end{equation}
where now
\begin{eqnarray}
I_n(G,\Sigma) &=&  \Bigg\{ \text{Tr}\log\left[  i \omega_n\delta_{\alpha\beta} + i\frac{a}{T}\sigma^z_{\alpha\beta} + \Sigma_{\alpha\beta}(\omega_n) \right]  \nonumber\\
&& \qquad\quad - \Sigma_{\alpha\beta}(\omega_n) G_{\beta\alpha}(\omega_n) + (-1)^{\alpha+\beta}  \frac{m^2}{2}  G_{\alpha\beta}(\omega_n)G_{\beta\alpha}(\omega_n)  \Bigg\}
\nonumber\\
&=& \Bigg\{ \text{Tr}\log\left[ \left( i \omega_n\sigma^z_{\alpha\gamma}  + i\frac{a}{T}\delta_{\alpha\gamma} + \tilde{\Sigma}_{\alpha\gamma}(\omega_n) \right) \sigma^z_{\gamma\beta} \right] \nonumber\\
&& \qquad\quad - (-1)^{\alpha+\beta} \tilde{\Sigma}_{\alpha\beta}(\omega_n) \tilde{G}_{\beta\alpha}(\omega_n) +\frac{m^2}{2}  \tilde{G}_{\alpha\beta}(\omega_n)\tilde{G}_{\beta\alpha}(\omega_n) \Bigg\}\,, 
\end{eqnarray}
In this expression repeated indices are summed over, the trace is taken over the replica indices, and in the second equality we have defined $\tilde{G}=G\sigma^z$ and $\tilde{\Sigma}=\Sigma\sigma^z$. This may be rewritten as 
\begin{eqnarray}\label{Iinta}
I(G,\Sigma) &=& \text{Tr}\log\left[  \left( i \sigma^z_{\alpha\gamma}\delta^{nn'}\omega_n + i\frac{a}{T}\delta^{nn'}_{\alpha\gamma} + \tilde{\Sigma}_{\alpha\gamma}^{nn'}\right) \sigma^z_{\gamma\beta}  \right] - \text{Tr}\left[ \left( (-1)^{\alpha+\gamma} \tilde{\Sigma}_{\alpha\gamma}^{np} -\frac{m^2}{2}  \tilde{G}_{\alpha\gamma}^{np} \right) \tilde{G}_{\gamma\beta}^{pn'} \right] , \label{Imaj} \nonumber \\
\end{eqnarray}
with the traces now taken over both replica and (odd $n$) frequency spaces. In addition, we have introduced the frequency space indices $n,n',p$ and defined the matrix $\delta^{nn'}_{\alpha\beta}\equiv\delta^{nn'}\delta_{\alpha\beta}$.  

\subsection{Saddle point approximation}

The overall factor of $N$ appearing in (\ref{SFFeff1}) makes the SFF well suited to a saddle point analysis in the large $N$ limit.  From \eqref{Icomplexa} we find the $(\tilde{G},\tilde{\Sigma})$ equations of motion,
\begin{align}
    \frac{\delta I}{\delta \tilde{\Sigma}_{\alpha\beta}^{nn'}} = 0 \quad &\Rightarrow \quad  \sigma^z_{\alpha\rho} \left[ i\sigma^z_{\rho\gamma}\delta^{np}\omega_n + i\frac{a}{T}\delta_{\rho\gamma}^{np} + \tilde{\Sigma}_{\rho\gamma}^{np}\right] \sigma^z_{\gamma \nu}
    (-1)^{\nu+\beta} \tilde{G}_{\nu\beta}^{pn'} = \delta_{\alpha\beta}^{nn'}, \\
    \frac{\delta I}{\delta \tilde{G}_{\alpha\beta}^{nn'}} = 0 \quad &\Rightarrow \quad \tilde{\Sigma}_{\alpha\beta}^{nn'} = (-1)^{\alpha+\beta} m^2 \tilde{G}_{\alpha\beta}^{nn'},
\end{align}
which can be written as a single equation of motion for $\tilde{\Sigma}$,
\begin{align}\label{offdiagCaSaa}
     \left( i\sigma^z\omega + i\frac{a}{T} +\tilde{\Sigma} \right) \tilde{\Sigma} =m^2,
\end{align}
suppressing the frequency and replica indices.  Similarly, we can rewrite the on-shell action (\ref{Iinta}),
\begin{eqnarray}   \label{ISigmaTilde}
I(\tilde{\Sigma}) &=& \text{Tr}\log\left[  (i \sigma^z \omega + i\frac{a}{T} + \tilde{\Sigma}) \sigma^z \right] - \frac{1}{2} \text{Tr} \tilde{\Sigma}^2\,, \label{Iga}
\end{eqnarray}
Note that without the $\sigma^z \omega$ term, the action (\ref{ISigmaTilde}) is invariant under general unitary transformations $\tilde{\Sigma}\rightarrow U^\dag\tilde{\Sigma}U$. In the presence of the $\sigma^z \omega$ term, this symmetry is explicitly broken to transformations, $\tilde{\Sigma}\rightarrow \tilde{U}^\dag\tilde{\Sigma}\tilde{U}$ for unitaries $\tilde{U}$ satisfying $\tilde{U} \sigma^z \omega \tilde{U}^\dag=\sigma^z \omega$, where we recall that $\sigma^z \omega$ is actually a matrix in replica and frequency space given with matrix elements $(\sigma^z \omega)_{\alpha\beta}^{nn'} = \sigma^z_{\alpha\beta} \delta^{nn'}\omega_n$. 
The breaking of this symmetry and the associated Goldstone zero modes play a crucial role in giving rise to the exponential ramp exhibited by the SFF in this system \cite{Liao:2021ofk}. \\ \\
Solutions to (\ref{offdiagCaSaa}) are diagonal in both replica and frequency indices, and are given by
\begin{align}
\tilde{\Sigma}_{\alpha\beta}^{nn'} = \frac{1}{2} \left[ -i\left(\xi_\alpha \omega_n +\frac{a}{T}\right) + c_\alpha^n \sqrt{4m^2-\left(\xi_\alpha \omega_n +\frac{a}{T}\right)^2}\right]\delta^{nn'}_{\alpha\beta},
\end{align}
where $\xi_1=1, \ \xi_2=-1$ and the constants $c_\alpha^n = \pm1$ can be chosen independently for each replica and frequency index. Imposing the condition that for large frequencies our solution must vanish \cite{Altland:2000fj} we obtain the following solutions
\begin{align}\label{gsolnsa}
\mathcal{S}_{\alpha\beta}^{nn'} = \begin{cases} \frac{1}{2} \left[ -i(\xi_\alpha x_n+b) + c_\alpha^n \sqrt{4-(\xi_\alpha x_n +b)^2}\right]\delta^{nn'}_{\alpha\beta}  ,&  |\xi_\alpha x_n+b|\leq 2,  \\
    \frac{1}{2} \left[ -i(\xi_\alpha x_n +b) + i\text{sgn}(\xi_\alpha x_n +b) \sqrt{(\xi_\alpha x_n +b)^2-4}\right]\delta^{nn'}_{\alpha\beta}, &  |\xi_\alpha x_n +b| > 2.
\end{cases}
\end{align}
in terms of the dimensionless quantities $\mathcal{S}\equiv \frac{1}{m}\tilde{\Sigma}$, $x_n\equiv\frac{\omega_n}{m}=\frac{n \pi}{mT}$ and $b\equiv\frac{a}{mT}$.  Evidently, there are many saddles, one for each value of the constants $c_\alpha^n$ for frequencies in the range $|\xi_\alpha x_n+b|\leq 2$. It remains to check which of these contribute to the SFF at large $N$.\\

\noindent
In what follows,
we work with a form of on-shell action (\ref{Iga}) that has been regularized by subtracting off the action for the free theory $\log(i\omega)$, to obtain
\begin{eqnarray}\label{Igr}
I(\mathcal{S}) &=& \text{Tr}\log\left[  \left( \sigma^z  + \frac{b}{x} - \frac{i}{x}\mathcal{S} \right) \sigma^z \right] - \frac{1}{2} \text{Tr} \mathcal{S}^2\,,  \label{Igra}
\end{eqnarray}
where, from here on, we suppress the frequency dependence of $x_n$. To see which saddle points solutions contribute to the SFF, we must analyse fluctuations $\delta\mathcal{S}^{nn'}_{\alpha\beta}$ about the various saddles $\mathcal{S}^{nn'}_{\alpha\beta}$. To this end, consider the variation of the on-shell action (\ref{Igra}) under $\mathcal{S}\rightarrow \mathcal{S}+\delta \mathcal{S}$.  After simplifying the expression by using the equation of motion (\ref{offdiagCaSaa}) we find that
\begin{align}\label{fluctactiona}
    \delta I (\mathcal{S}) &= -\frac{1}{2} \sum_{\alpha,\beta,n,n'} M^{nn'}_{\alpha\beta}\delta\mathcal{S}^{nn'}_{\alpha\beta}  \delta\mathcal{S}^{n'n}_{\beta\alpha},
\end{align}
with the mass matrix
\begin{align}
    M^{nn'}_{\alpha\beta} \equiv 1 + \mathcal{S}^{nn}_{\alpha\alpha} \mathcal{S}^{n'n'}_{\beta\beta}.
\end{align}
Note that here $\mathcal{S}^{nn}_{\alpha\alpha}$ denotes a single diagonal component of the matrix $\mathcal{S}$ and not a trace ({\it i.e.} no summation over repeated indices is implied in the above expression).  Expressed in the above terms, and normalising by dividing by $\overline{Z(0)}^{2}$, in the saddle point approximation,
\begin{equation}
g(T) = \mathcal{N} \sum_{\textnormal{saddles}} e^{N I( \mathcal{S}_{cl})  } \prod_{\alpha, \beta}  \int \left(\prod_{n,n' \ \textnormal{odd}} \mathscr{D} \delta \mathcal{S}_{\alpha \beta}^{n n'} \right) e^{N \delta I(\mathcal{S}_{cl} ) }\,.
\end{equation}

\subsection{The contributing saddles}

For each frequency there are four possibilities, dictated by the values of $|\xi_1 x_n + b|$ and $|\xi_2 x_n + b|$.  If both are less (greater) than $2$ we refer to this region as $A$ ($D$).  Regions $B$ and $C$ cover the other two possibilities where either $|\xi_1 x_n + b| \leq 2$ or $|\xi_2 x_n + b| \leq 2$ respectively.  \\ \\
In Appendix \ref{MassMatrixAppendix} we verify that the mass matrices always have a non-negative real part in all four regions, ensuring that all saddles do contribute.  Moreover, the real part of the mass matrices are all strictly positive except when $\alpha \neq \beta$, $n' = -n$ inside region $A$, where the vanishing mass matrix signals the presence of zero modes which dominate the contribution to the fluctuations.  The region $A$ is, however, only present if $|b| = |\frac{a}{m T}| < 2$.  For a given gauge field, this sets a natural scale for the disorder strength necessary for zero modes to appear.  \\ \\
Furthermore, note that the mass matrices satisfy the following complex conjugation relation
\begin{equation}
    \label{massConjugate}
    \left( M^{nn'}_{\alpha\beta}\big|_{c_1^{n}=c_2^{n'}=\pm1} \right)^*  =  M^{nn'}_{\alpha\beta}\big|_{c_1^{n}=c_2^{n'}=\mp1}\,,    
\end{equation}
in each of the four regions, a fact that will be important in demonstrating that we obtain a real expression for the SFF. Using (\ref{Igr}) we define $I_{J}(x) \equiv I(\mathcal{S}^{(J)})$ where $J=A,B,C,D$, so the classical action takes the form,
\begin{align}\label{SFFbl0Ta}
I(S_{cl})=&  2  \Bigg[   \sum_{\substack
  {n\pi>0\\n \ \text{odd}}}^{2mT-|a|} I_A^{(c,\tilde{c})}(x_n) +  \frac{1}{2} \sum_{\substack{ n\pi> 2mT-|a|\\n \ \text{odd}}}^{2mT+|a|}  \left(I_{C}^{(c)}(\text{sgn}(a) x) + I_{C}^{(\tilde{c})}(\text{sgn}(a) x) \right) +\!\!\!\!\!\!\!
\sum_{\substack
  { n\pi> 2mT+|a|\\n \ \text{odd}}}^{\infty}\!\!\!\!\!\! I_D(x_n) \Bigg],
\end{align}
for strong disorder $|a|/T < 2m$, while for weak disorder $|a|/T > 2m$,
\begin{align}\label{SFFbg0Ta}
    I(S_{cl})=&  2 \Bigg[ \sum_{\substack
  {n\pi>0\\n \ \text{odd}}}^{|a|-2mT} I_D(x_n) + \frac{1}{2}\sum_{\substack
  { n\pi>|a|-2mT\\n \ \text{odd}}}^{|a|+2mT}  \left(I_{C}^{(c)}(\text{sgn}(a) x) + I_{C}^{(\tilde{c})}(\text{sgn}(a) x) \right)  + \!\!\!\!\!\!\!\sum_{\substack
  { n\pi> 2mT+|a| \\n \ \text{odd}}}^{\infty}\!\!\!\!\!\! I_D(x_n) \Bigg] .
\end{align}
where the sums run over positive values of $x$ only and we have used the symmetries of $I_D(x)=I_D(-x)$, $I^{(c)}_B(x)=I^{(c)}_C(-x)$ and $I^{(c,\tilde{c})}_A(x)=I^{(\tilde{c},c)}_A(-x)$. Note that the factors of $I_A$, $I_{B}$ and $I_C$ all depend on the values $c\equiv c_1^n, \tilde{c}\equiv c_2^n$ labelling the various saddles while $I_D$ is the same for all saddle points.  The sum over saddle points then involves a sum over all allowed choices of $c_\alpha^n$. 
To proceed we will evaluate these expressions in the early- and late-time limits.  

\subsection{SFF for $mT\ll1$}

In the early-time limit, all the contributions to the Matsubara sums in (\ref{SFFbl0Ta}),(\ref{SFFbg0Ta}) come from $I_D$, unless there is an odd integer $n$ within $2mT$ of $|a|/\pi$.\footnote{In this case there is a single-frequency contribution from $I_C$ constituting four saddles points, see Appendix \ref{appendix:D}.} When $a$ is not too close to being an odd multiple of $\pi$ we may simply neglect this contribution.  Since $I_D$ does not involve the constants $c_\alpha^n$ there is a single saddle in this limit and we find that
\begin{eqnarray}
   e^{N I( \mathcal{S}_{cl})} &=& \exp\Bigg\{ 2N \sum_{n>0,\text{odd}} I_D(x_n) \Bigg\} \nonumber\\
     &\approx & \prod_{n>0,\text{odd}} \left[\frac{(n\pi-a)(n\pi+a)}{n^2\pi^2}\right]^{2N} \exp\left\{-N m^2 T^2 \sum_{n>0,\text{odd}}\frac{2 \left(n^2\pi^2+a^2\right)}{(n\pi-a )^2 (n\pi+a )^2} \right\} \nonumber\\
    &= & \cos\left(\frac{a}{2}\right)^{2N} \exp\Bigg\{ -\frac{N m^2T^2}{4} \sec ^2\left(\frac{a}{2}\right) \Bigg\},
\end{eqnarray}
where, in the second equality, we have neglected $\mathcal{O}(m^4T^4)$ terms in the exponent. The fluctuation integral does not involve zero modes and we may approximate it as a constant at early times. After normalising by dividing by $g(0)=|Z(0)|^2$, we obtain the following expression for the spectral form factor
\begin{align}\label{mTll1SFFaa}
    g(T) = & \exp\left\{ -\frac{N m^2T^2}{4} \sec ^2\left(\frac{a}{2}\right) \right\}.
\end{align}
 In the limit $a\rightarrow0$ the argument of the exponent reduces to $-Nm^2T^2/4$, in agreement with the Majorana result in \cite{Winer:2020mdc}\footnote{Note that our result is a factor of 2 larger, arising from the fact that complex fermions have twice as many degrees of freedom as Majorana fermions.}. We note also, in passing, that \eqref{mTll1SFFaa} is periodic in $a$ with period $2\pi$ ensuring that the early-time SFF is indeed gauge invariant under large gauge transformations {\it i.e.} transformations for which $a\rightarrow a+2\pi k$, $k\in \textbf{Z}$.
\\ \\
The exponent of (\ref{mTll1SFFaa}) blows up to negative infinity as $a\to\pi$. However, recall that our result is valid only up to values of $mT$ for which $a$ is a distance $2mT$ away from $\pi$. We refer the reader to Appendix \ref{appendix:C} where we plot the analytic results up to $\mathcal{O}(m^8T^8)$ for various values of $a$ and Appendix \ref{appendix:D} for a more careful treatment in the region $\pi-a<2mT\ll1$.

\subsection{SFF for $mT\gg1$}

At late times (and assuming strong disorder), we use the expression (\ref{SFFbl0Ta}) for the SFF and  continue $x_n$ to a continuous variable so that the Matsubara sums are approximated by integrals. This gives 
\begin{eqnarray}\label{largemTa}
    S_{cl} &=&   \frac{mTN}{\pi}  \Bigg[\int_{0}^{2-|b|}dx\, I_A^{(c,\tilde{c})}(x) + \frac{1}{2}\int_{2-|b|}^{2+|b|}dx \left(I_{C}^{(c)}(\text{sgn}(b) x) + I_{C}^{(\tilde{c})}(\text{sgn}(b) x) \right)\nonumber\\  
    &+& \int_{2+|b|}^{\infty}dx\, I_D(x) \Bigg]\,.
\end{eqnarray}
Summing over saddles labelled by all possible configurations of $(c,\tilde{c})$, results in a vanishing SFF (see Appendix \ref{appendix:F}). To see a nonvanishing result we need to consider fluctuations about these saddles. Of particular interest to us in the expression (\ref{largemTa}) is the inclusion of the region $A$ where zero modes are possible. As we will see, the effect of these zero modes is to give rise to an exponential ramp at late times. Moreover, the presence of the gauge field $a$ sets a natural scale $\lambda \equiv \frac{|a|}{2m}$ for the disorder in order for zero modes to be supported. To elaborate, let's rewrite the expression for the SFF as,
\begin{align}
    g(T) &= \frac{\int \mathscr{D}\mathcal{S} e^{N I(\mathcal{S})}}{\int \mathscr{D}\mathcal{S} e^{ -\frac{N}{2} \text{Tr}\mathcal{S}^2 }},
\end{align}
where we now normalise by dividing by the free theory \cite{Winer:2020mdc, Liao:2020lac}
\begin{align}
    \int \mathscr{D}\mathcal{S} e^{ -\frac{N}{2} \text{Tr}\mathcal{S}^2 } = \prod_{\alpha,\beta,n,n'} \sqrt{\frac{2\pi}{N }}.
\end{align}
We can then write
\begin{align}\label{gnorm}
    g(T)=\sum_{c,\tilde{c}} e^{N I(\mathcal{S})} F(T),
\end{align}
using the large $N$ saddle point approximation up to quadratic fluctuations and summing over all saddles labelled by $c\equiv c_\alpha^n ,\tilde{c}=c_\beta^{n'}$. The quadratic fluctuation path integral, containing the normalisation, is given by 
\begin{align}\label{flucPIa}
    F(T) 
    &= \prod_{\alpha,\beta,n,n'} \sqrt{\frac{N}{2\pi}} \int \mathscr{D}\delta \mathcal{S}_{\alpha\beta}^{nn'} \exp\left\{ - \frac{N}{2} M_{\alpha\beta}^{nn'} \delta\mathcal{S}_{\alpha\beta}^{nn'}\delta\mathcal{S}_{\beta\alpha}^{n'n}  \right\}.
\end{align}
using (\ref{fluctactiona}). Note that the mass matrix is dependent on the particular saddle point solution  considered.  We have already verified that $M_{\alpha\beta}^{nn'}\geq0$ for all saddles and frequency regions.  From the cases where $M_{\alpha\beta}^{nn'}>0$, we expect a finite contribution to the path integral, provided that the products over frequencies converge.  We also have zero modes in region $A$ where $M_{\alpha\beta}^{nn'}=0$ for some saddles. \\

\noindent
Let us first consider massive modes and their fluctuations. For these modes $M_{\alpha\beta}^{nn'}>0$ and the quadratic fluctuation integral is easily carried out in each case to obtain
\begin{align}\label{massivefluct}
      F(T) = \prod_{\alpha,\beta,n,n'} \sqrt{\frac{N }{2\pi}} \sqrt{\frac{2\pi}{N M_{\alpha\beta}^{nn'}}}  = \exp\left\{ -\frac{1}{2} \sum_{\alpha,\beta,n,n'}\log M_{\alpha\beta}^{nn'} \right\}\,.
 \end{align}
Note that the contribution from these massive fluctuations is independent of $N$. While we will not evaluate it explicitly, we will verify that it is convergent and results in a real SFF. 
\\ \\
First, let's prove convergence.  The regions $A$, $B$ and $C$ are finite, so that the contribution from the massive modes from this regions is clearly finite. In the infinite region $D$, however, we must entertain the possibility of UV divergences. For very large frequencies $x_n,x_{n'}\gg1$, the mass matrix given explicitly in (\ref{DM}) can be approximated by
\begin{align}
    M_{\alpha\beta}^{nn'} &\approx 1 - \frac{4}{\xi_\alpha\xi_\beta x_n x_{n'}} \quad\Rightarrow\quad  \sum_{\alpha,\beta,n,n'} \log M_{\alpha\beta}^{nn'} \sim \sum_{\alpha,\beta,n,n'} \frac{1}{\xi_\alpha\xi_\beta x_n x_{n'}}.
\end{align}
The sum is not absolutely convergent, but by imposing any finite frequency cutoff we can ensure that cancellations between either $n'=-n$ or $\alpha\neq\beta$ are sufficient to deal with any UV divergences \cite{Winer:2020mdc}.
\\ \\  Second, the realness of the SFF follows from the complex conjugation relations of the mass matrices (\ref{massConjugate}), and that the classical action (\ref{largemTa}) exhibits the same complex conjugate relations.  Consequently, after summing over the saddles, one obtains a real answer.  
\\ \\
We now turn our attention to the zero modes.  As discussed, these occur in the region $A$ where mass matrices with $\alpha\neq\beta$ and $n'=-n$ vanish when we are considering a saddle with $c_1^{n}=-c_2^{-n}$. Due to the vanishing of the mass matrix we need to compute the contribution of the zero modes differently. The zero modes are Goldstone modes that have their origin in the explicit breaking of the symmetry of the equations of motion (\ref{offdiagCaSaa}) under general unitary transformations $\tilde{\Sigma}\rightarrow U^\dag\tilde{\Sigma}U$ by the $\sigma^z \omega$ term.  The full set of zero modes are obtained by performing unitary transformations with the residual symmetry and summing over solutions that are not related in this way.  Zero modes that are related by such unitary transformations are degenerate.  To be explicit, for each non-degenerate zero mode pair, from the normalisation with respect to the free contribution, we obtain a factor of $\frac{N}{2\pi}$.  This should be multiplied by the volume of the factor group representing the residual symmetry \cite{Kamenev:1999zz}.  This volume does not depend on $N$.  Since the massive modes also give a contribution that is $N$-independent, the leading large $N$ SFF is obtained by considering the saddle points that maximize the number of non-degenerate zero modes. The classical action for these saddle points vanishes and the total contribution to the SFF for such saddles is given by
\begin{eqnarray}\label{zeroramp}
   2N^{2\left(2mT - |a| \right)/\pi} \left( \prod_{|x_n| <2-\frac{|a|}{mT}}  \frac{\mathcal{V}^n_{12} \mathcal{V}_{21}^n}{2\pi} \right) \exp\left\{ -\frac{1}{2} \sum_{\alpha,\beta,n,n'}^{'}\log M_{\alpha\beta}^{nn'} \right\} ,
\end{eqnarray}
where the summation $\sum_{\alpha,\beta,n,n'}^{'}$ excludes the zero modes.  The first factor above is the leading large-$N$ contribution from maximising the number of non-degenerate zero modes; the second is the product of the $N$-independent factor group volumes and the third is the contribution from the massive modes for the saddle. Note that, though this expression for the ramp is not gauge-invariant under large gauge transformations $a\rightarrow a+2\pi k$, we  expect gauge-invariance to be restored when higher order fluctuations are taken into account.
\\ \\
Up to quadratic order in fluctuations, all $N$ dependence is contained in the prefactor $N^{2\left(2mT - |a| \right)/\pi}$, from which we observe the exponential ramp typical of noninteracting disordered systems \cite{Winer:2020mdc}. Note that both $m T$ and $\frac{|a|}{T} T$ are important and we require strong disorder in order for the ramp to be present. 
We expect the contribution from the remaining modes contained in the exponential term to become important at times $mT\gg N$, where the soft modes combine with the zero mode fluctuations to produce the late time plateau chacteristic of the SFF \cite{Winer:2020mdc}. In addition, we expect that the coefficient in the $N$ dependent ramp may receive corrections from higher order fluctuations than the quadratic ones we have considered here \cite{Liao:2020lac}. Note also that we have neglected saddle point solutions which break time translation invariance, which may slightly alter the results obtained. \\ \\
For $mT\ll1$ on the other hand, we had no contribution from region $A$. Consequently, there are no zero mode fluctuations to consider. The remaining massive mode fluctuations are subleading in $N$ and so may be neglected.  Thus our results for the $mT\ll1$ SFF slope in (\ref{mTll1SFFaa}) and (\ref{mTll1SFFa}) are in harmony with the late time results.

\section{Discussion}

In this paper, we have studied the properties of random mass fermions in one dimension with twisted boundary conditions. The introduction of random masses (for the fermions) furnishes a promising toy model to study aspects of an $AdS_2$/gauged-SYK correspondence. This is due to two main features that this model inherits: i. the random $q$-body interaction characteristic of  SYK$_q$ models and ii. the gauge field term that has been shown \cite{Dunne:1996yb} to endow the $0+1$ dimensional model with features of $2+1$-dimensional Chern-Simons theory, as well as the solvability of both. As discussed in the introduction, our original goal was to construct a gauged SYK$_{q}$ model with an eye on the canonical $q=4$ case largely because of its known holographic connection to JT gravity. However as we hope to have convinced any reader to have made it to this point, the $q=2$ (quadratic) model is also rather rich and certainly merits the detour.\\

\noindent
As emphasized in the main text, one may naively expect that coupling the fermions to a gauge field has the same effect as turning on a complex chemical potential.  However, the gauge field transforms differently to a chemical potential. An important check is that all physical quantities be invariant under large gauge transformations.  Though  the complex SYK model (for large $N$) is solvable, it is not immediately obvious that the introduction of a gauge field (or equivalently applying twisted boundary conditions) will retain this crucial property. Happily, we found that it does, at least in this quadratic case.\\

\noindent
As we have noted, the complex SYK$_2$ model with chemical potential has been considered in \cite{Liao:2020lac}, largely using different methods to the path integral approach utilised here and in \cite{Winer:2020mdc}. Their model is equivalent to ours under the replacement
of the chemical potential $\mu$ by the gauge field $iA(\tau)$. What distinguishes our model is that in periodic \textit{real time} with period $T$, relevant for the calculation of the SFF, the constant gauge configuration\footnote{Recall that in Euclidean time, the constant gauge configuration is $A(\tau)=a/\beta$.} becomes $A(t)=a/T$. 
Since $T$ is the time argument of the SFF, the external gauge field in our model therefore introduces additional ``time''  dependence into the SFF calculation that is absent in the case of a constant chemical potential. Indeed, our SFF results are not equivalent to those of \cite{Liao:2020lac} under the analytic continuation $\mu\rightarrow  ia/T$. Under this replacement, the result for the SFF slope\footnote{We have correctly normalised their result and taken the limit $T\ll0$ and $m=1$} in \cite{Liao:2020lac},
\begin{align}
    g(T) = \exp\left\{ - \frac{N T^2}{2} \cos (\mu T) \right\} \rightarrow \exp\left\{ - \frac{N T^2}{2} \cosh (a) \right\} ,
\end{align}
has qualitatively different behaviour and is not invariant under large gauge transformations.
\\

\noindent
The primary foci of this article were the out-of-time-ordered correlator and spectral form factor of this model, using a quenched and annealed averaging protocol respectively. From a technical viewpoint many of the established tools for solving the SYK$_2$ model can be readily adapted to the present case.  In particular, the fully dressed two-point correlation function can be computed exactly at large $N$.  When transforming to real time and Wick rotating the gauge field, the gauge field gives rise to an overall phase factor for the two-point function. When considering expressions involving a balance of the Green's function and its conjugate, these phases may cancel.  In particular, the out-of-time-order correlator is unaffected with no exponential growth for the out-of-time-order correlator, confirming that the SYK$_{2}$ model's integrablity persists at early times even with the gauge field turned on.  \\

\noindent 
The spectral form factor is more interesting. Here we observe two noteworthy features introduced by the gauge field.  The first is an enhancement of the early time slope.  
The second is the introduction of an explicit scale for strong disorder necessary for the presence of zero modes. These zero modes give rise to an exponential ramp at late times, consistent with existing results for the quadratic SYK model.  \\

\noindent
Our calculation results in an SFF slope controlled by the parameter $a$, which can be thought of as taking values in the interval $[0,2\pi)$ due to the invariance under large gauge transformations. For $a=0$ we obtain the result of the ungauged theory studied in \cite{Winer:2020mdc}. As we increase $a$ from 0 to $\pi$, the slope decays faster and faster until the exponent diverges to $-\infty$ for $a=\pi$.  
\\

\noindent
The fact that we may use the parameter $a$ to tune the decay of the slope presents an interesting possibility. In a quantum chaotic system there is a timescale which signals the onset of RMT universality; the Thouless time \cite{Nosaka:2018iat}. From the perspective of the SFF, it is the time at which the ramp begins, but identifying it for a given model is nontrivial, because the beginning of the ramp is usually obscured by the slope. In order to find the Thouless time, some method is then required to isolate the ramp from the slope. 
\\

\noindent
In a Gaussian Unitary Ensemble (GUE) this can be achieved by simply calculating the connected and disconnected pieces of the SFF separately; the disconnected part contains the slope while the connected part contains the ramp. However, for the SYK$_4$ model, this technique  fails due to the fact that the connected piece contains its own slope \cite{Cotler:2017jue}. Various ways of dealing with this have been proposed \cite{Nosaka:2018iat} involving variations of the SFF. The ``Gaussian-filtered SFF'' \cite{Gharibyan:2018jrp} eliminates the slope by softening the hard edge of the spectral density from which it arises, thus allowing for the ramp to be visible at earlier times. The ``connected unfolded SFF'' \cite{Garcia-Garcia:2018ruf} is the SFF computed using the unfolded spectrum and removing the disconnected piece.
\\

\noindent
Our SFF results show that for SYK$_2$ model, the twisting of the fermion boundary conditions doesn't significantly affect the large $T$ exponential ramp, while at small $T$ we can kill off the slope by taking $a\rightarrow\pi$. If it were the case that taking $a\rightarrow\pi$ had a similar effect on the slope of the \textit{``gauged'' complex SYK}$_4$ model, while leaving the ramp invariant, then this limit could be used as another method of controlling the slope independently of the ramp, for the purposes of identifying the Thouless time. Further analysis of the gauged SYK$_4$ model would be necessary in order to make precise statements in this direction.
\\

\noindent
We briefly note the different mechanisms underlying the linear and exponential ramps in the Majorana SYK$_4$ model and the (Majorana and complex) SYK$_2$ models, respectively. For the Majorana SYK$_4$ model, the contributions of the replica-diagonal saddles to the full SFF just give the disconnected SFF \cite{Saad:2018bqo}, which describes the slope. The ramp can only be seen by including the late time contributions of the replica-non-diagonal saddles, which spontaneously break time translation symmetry. In the SYK$_2$ cases considered here and in \cite{Liao:2020lac,Winer:2020mdc}, we see that the ramp has different origins: considering only replica-diagonal solutions still leads to symmetry breaking and an exponential ramp, which is not present in the disconnected SFF\footnote{The replica-non-diagonal solutions then appear to play less of a role in these models.}.
\\

\noindent
Specifically, in the Majorana SYK$_2$ case studied in \cite{Winer:2020mdc}, the $SU(2)$ conjugation symmetry of the two-replica action is \textit{spontaneously} broken by the form of the diagonal saddle point solutions above a critical frequency. The associated Goldstone zero modes give rise to an exponential ramp. In the complex SYK$_2$ case studied here and in \cite{Liao:2020lac}, the 
the two-replica action contains a term which \textit{explicitly} breaks the unitary conjugation symmetry that it would otherwise possess. A subset of the associated Goldstone soft modes are indeed zero modes (with vanishing action) and these give rise to an exponential ramp. 
\\

\noindent
In our model, we found that the presence of the gauge field $a$ sets a natural timescale $\lambda \equiv \frac{|a|}{2m}$ in order for zero modes to be supported. However, as this scale is not gauge invariant, we expect that higher order corrections (beyond the quadratic fluctuations considered in this thesis) will have an effect.
\\

\noindent
In summary, our results show that coupling the SYK$_2$ model to a ``topological'' gauge field gives rise to a solvable model that retains the non-chaos properties of the former but which exhibits some interesting new features. This is a promising start for further development of this direction, specifically a gauged SYK$_{q}$ model. Evidently, the introduction of twisted boundary conditions with SYK$_q$ models do not present systems that are technically much more difficult than the SYK$_q$ models themselves. Of course, the obvious test of this statement will be the maximally chaotic $q=4$ model.
\\

\noindent
In particular, a natural follow up to our work would be to study a twisted SYK$_2$ model perturbed by an SYK$_4$ interaction. In this case, with the ``gauge field'' turned off we obtain a complex version of the mass-deformed (Majorana) SYK model studied in \cite{Garcia-Garcia:2018ruf,Nosaka:2018iat}, which exhibits a chaotic-integrable transition. It would then be interesting to see whether this transition is affected by the presence of a nontrivial twist of the fermion boundary conditions. In \cite{Bhattacharya:2017vaz} it was found that the Lyapunov exponent of the complex SYK$_q$ model in the $q\rightarrow\infty$ limit is suppressed by a nonzero chemical potential, with the chemical potential having an exponentially large dominance over the $q$-body interaction coupling strength in determining the chaos behaviour of the system. In light of this it would be interesting to see whether a nonzero twisting parameter has this same effect of pushing the mass-deformed complex SYK model into an integrable phase for all values of the SYK$_4$ coupling, from the perspective of spectral statistics (the SFF), as opposed to the OTOC. 
\noindent
An analytic treatment of the SFF for the mass deformed SYK model is still outstanding\footnote{A qualitative discussion from an analytic perspective is given in \cite{Winer:2020mdc}.}, so it is likely that the prospective calculations discussed above would have to be performed numerically. However, steps in this direction have already been taken for the complex SYK$_2$ model when perturbed by quartic, non-SYK$_4$ interactions. Namely, the SFF of the SYK$_2$ model perturbed by a \textit{non-random} 4-fermi interaction term \cite{Liao:2021ofk, Lau:2020qnl} and by a random 4-fermi SYK$_2^2$ interaction (the square of an SYK$_2$ term) \cite{Liao:2021uwx} have been studied. In both cases the solution for the SYK$_2$ SFF serves as the conceptual and analytical solution about which the full solutions are constructed as perturbations, with the result that the quartic interactions induce a mass for the SYK$_2$ soft modes which would otherwise give rise to the exponential ramp\footnote{While the model in \cite{Liao:2021ofk} is chaotic, the model in \cite{Liao:2021uwx} has chaotic and integrable regimes. In the integrable regime no mass is induced for the soft modes and the exponential ramp survives.}. The exponential ramp is then suppressed, a necessary prerequisite for the emergence of RMT statistics.
\\

\noindent
Another natural continuation of our work would then be to repeat our calculation for the ``gauged'' SYK$_2$ model perturbed by such non-SYK four-fermi interactions, and to see if this has an effect on the induced mass which is responsible for the chaotic-integrable transition.  As we have emphasised repeatedly, the effect of the external gauge field is equivalent to introducing twisted boundary conditions for the theory which, at least based on our results here, would impact quantities computed by utilising several replicas.  
\\

\noindent
Finally, we circle back to our initial motivation and point out that a gauged SYK model may have a critical role to play in the formulation of a more conventional weakly-coupled, small curvature gravity dual. We hope that our work will stimulate further investigations of these topics in the future.

\acknowledgments

We would like to thank Jonathan Shock for his collaboration during earlier stages of this work, as well as his invaluable Mathematica expertise throughout. We would also like to extend our thanks to the analymous referee whose detailed constructive comments and observations regarding the equivalence of the external gauge field to a twisting of the fermion boundary conditions led to a much better understanding of the problem. JM acknowledges support from the “Quantum Technologies for Sustainable Development” grant from the National Institute for Theoretical and Computational Sciences of South Africa (NITHECS) and a Simons Associateship at the Abdus Salam International Center for Theoretical Physics, Trieste where some of this work was carried out. RPS is supported by a graduate fellowship from the National Institute for Theoretical and Computational Sciences and by the Shuttleworth Postgraduate Scholarship Programme.  H.J.R.vZ is supported by the ``Quantum Technologies for Sustainable Devlopment" grant from the National Institute for Theoretical and Computational Sciences (NITHECS).


\bibliographystyle{JHEP}
\bibliography{sykrefs.bib}

\newpage
\appendix

\section{Derivation of the 2-replica SFF}\label{appendix:A}

Defining
\begin{align}
    I_\alpha = -\int_0^T dt \sum_{i=1}^N \psi_{i_\alpha}^\dag \frac{d}{dt} \psi_{i_\alpha},
\end{align}
we will explicitly perform the disorder average using the probability distributions (\ref{Pmij}), to obtain the 2-replica SFF. Note that we take each replica to have the same gauge field configuration, so the Chern-Simons term cancels out.
\begin{eqnarray}
&& \overline{Z(iT) Z(iT)^*} \nonumber\\
& = & \int \left[ \prod_{\alpha=1}^2 \prod_{i_\alpha=1}^{N} \mathscr{D}\psi_{i_\alpha}^\dag \mathscr{D}\psi_{i_\alpha}\right] e^{I_1+I_2} \left(\sqrt{\frac{N}{2\pi m^2}} \right)^N \left(\sqrt{\frac{N}{\pi m^2}} \right)^{N(N-1)} \exp\left\{ i\frac{a}{T} \int dt \left( \psi^\dag_{i_1}\psi_{i_1} -  \psi^\dag_{i_2}\psi_{i_2} \right) \right\} \nonumber\\
&& \prod_{i} \int d m_{R, i i} \exp\left\{ -i \int dt \ m_{R,ii} (\psi_{i_1}^\dag \psi_{i_1}-\psi_{i_2}^\dag \psi_{i_2}) \right\} e^{-\frac{N}{2m^2} m_{R,i i}^2} \nonumber\\
&& \prod_{i< j} \int d m_{R,ij} \exp\left\{ -i\int dt \ m_{R,ij} (\psi_{i_1}^\dag \psi_{j_1} + \psi_{j_1}^\dag \psi_{i_1} - \psi_{j_2}^\dag \psi_{i_2} - \psi_{i_2}^\dag \psi_{j_2}) \right\} e^{-\frac{N}{m^2} m_{R,i j}^2} \nonumber  \\
& & \prod_{i< j} \int d m_{I,ij} \exp\left\{ \int dt \ m_{I,ij} (\psi_{i_1}^\dag \psi_{j_1} - \psi_{j_1}^\dag \psi_{i_1} + \psi_{j_2}^\dag \psi_{i_2} - \psi_{i_2}^\dag \psi_{j_2}) \right\} e^{-\frac{N}{m^2} m_{I,i j}^2} \nonumber 
\\ 
& = & \int \left[ \prod_{\alpha=1}^2 \prod_{i_\alpha=1}^{N} \mathscr{D}\psi_{i_\alpha}^\dag \mathscr{D}\psi_{i_\alpha}\right] e^{I_1+I_2} \exp\left\{ i\frac{a}{T} \int dt \left( \psi^\dag_{i_1}\psi_{i_1} -  \psi^\dag_{i_2}\psi_{i_2} \right) \right\}  \nonumber\\
&& \exp\left\{- \frac{m^2}{2N} \int dt dt' \sum_{i} (\psi_{i_1}^\dag \psi_{i_1}(t) - \psi_{i_2}^\dag \psi_{i_2}(t))
( \psi_{i_1}^\dag \psi_{i_1}(t') - \psi_{i_2}^\dag \psi_{i_2}(t') ) \right. \nonumber \nonumber\\
&& \left. - \frac{m^2}{4N} \int dt dt' \sum_{i<j} (\psi_{i_1}^\dag \psi_{j_1}(t) + \psi_{j_1}^\dag \psi_{i_1}(t) - \psi_{i_2}^\dag \psi_{j_2}(t) - \psi_{j_2}^\dag \psi_{i_2}(t)) \right. \nonumber\\
&& \left. \qquad\qquad\qquad\qquad\qquad  (\psi_{i_1}^\dag \psi_{j_1}(t') + \psi_{j_1}^\dag \psi_{i_1}(t') - \psi_{i_2}^\dag \psi_{j_2}(t') - \psi_{j_2}^\dag \psi_{i_2}(t')) \right.  \nonumber
\\
&& +\frac{m^2}{4N} \int dt dt' \sum_{i<j}
(\psi_{i_1}^\dag \psi_{j_1}(t) - \psi_{j_1}^\dag \psi_{i_1}(t) - \psi_{i_2}^\dag \psi_{j_2}(t) + \psi_{j_2}^\dag \psi_{i_2}(t)) \nonumber\\
&&  \qquad\qquad\qquad\qquad 
(\psi_{i_1}^\dag \psi_{j_1}(t') -  \psi_{j_1}^\dag \psi_{i_1}(t') - \psi_{i_2}^\dag \psi_{j_2}(t') + \psi_{j_2}^\dag \psi_{i_2}(t')) \Bigg\}  \nonumber  
\\ 
\\
& = & \int \left[ \prod_{\alpha=1}^2 \prod_{i_\alpha=1}^{N} \mathscr{D}\psi_{i_\alpha}^\dag \mathscr{D}\psi_{i_\alpha}\right] e^{I_1+I_2}  \exp\Bigg\{ - \frac{m^2}{2N} \int dt dt' \sum_{\alpha,\beta} \sum_{i,j} (-1)^{\alpha+\beta} \psi_{i_\alpha}^\dag \psi_{j_\alpha}(t)\psi_{j_\beta}^\dag \psi_{i_\beta}(t') \Bigg\} \nonumber \\ 
&&\exp\left\{ i\frac{a}{T} \int dt \left( \psi^\dag_{i_1}\psi_{i_1} -  \psi^\dag_{i_2}\psi_{i_2} \right) \right\}
\label{this2a}
\end{eqnarray}
We now define the collective, bilocal field
\begin{align}
    G_{\alpha\beta}(t',t) = \sum_{i_\alpha=1}^N \frac{1}{N}\psi^\dag_{i_\alpha}(t)\psi_{i_\beta}(t'),
\end{align}
which we introduce to the partition function via the functional delta function
\begin{align}
    & \delta\left( G_{\alpha \beta}(t',t)- \sum_{i=1}^N \frac{1}{N}\psi^\dag_{i_\alpha}(t)\psi_{i_\beta}(t') \right)\\
    &= \prod_{\alpha,\beta} \int \mathscr{D}\Sigma_{\alpha \beta} \exp\left\{ -N \Sigma_{\beta\alpha}(t,t') \left( G_{\alpha \beta}(t',t)- \sum_{i=1}^N \frac{1}{N}\psi^\dag_{i_\alpha}(t)\psi_{i_\beta}(t')\right) \right\}.
\end{align}
Note that we have introduced the auxilary field $\Sigma_{\beta\alpha}$ as a Lagrange multiplier. This allows us to rewrite the quartic term in (\ref{this2a}) as
\begin{align}
&\exp\left\{ -\frac{m^2}{2N} \int dt dt' \sum_{\alpha,\beta}\sum_{i,j} (-1)^{\alpha+\beta} \psi_{i_\alpha}^\dag \psi_{j_\alpha}(t)\psi_{j_\beta}^\dag \psi_{i_\beta}(t')  \right\} \nonumber
\\
= & \prod_{\alpha,\beta} \int \mathscr{D}\Sigma_{\alpha \beta} \mathscr{D}G_{\alpha \beta} \exp\Bigg\{ \int dt dt' \bigg[ (-1)^{\alpha+\beta} \frac{Nm^2}{2} G_{\alpha \beta}(t',t)G_{\beta\alpha}(t,t') \nonumber\\
& \qquad\qquad\qquad\qquad - N \Sigma_{\beta\alpha}(t,t') \left( G_{\alpha \beta}(t',t)  - \sum_{i=1}^N \frac{1}{N}\psi^\dag_{i_\alpha}(t)\psi_{i_\beta}(t')\right) \bigg] \Bigg\},
\end{align}
so that (\ref{this2a}) becomes
\begin{eqnarray}
&& \int \left[\prod_{\alpha,\beta} \mathscr{D}\Sigma_{\alpha \beta} \mathscr{D}G_{\alpha \beta} \right] \left[ \prod_\alpha \prod_{i_\alpha=1}^{N} \mathscr{D}\psi_{i_\alpha}^\dag \mathscr{D}\psi_{i_\alpha}\right]  \nonumber
\\
&&\exp\Bigg\{\int dt dt' \sum_{\alpha,\beta} \sum_{i} \psi_{i_\alpha}^\dag(t) \left[ -\delta(t-t')\left( \delta_{\alpha \beta}\frac{d}{dt} - i \sigma^z_{\alpha\beta} \frac{a}{T} \right) + \Sigma_{\beta\alpha}(t,t') \right] \psi_{i_\beta}(t') 
\nonumber\\
&& -N \int dt dt' \sum_{\alpha,\beta} \left[ \Sigma_{\beta\alpha}(t',t) G_{\alpha \beta}(t,t') - (-1)^{\alpha+\beta} \frac{m^2}{2} G_{\alpha \beta}(t,t')G_{\beta\alpha}(t',t) \right] \Bigg\},
\end{eqnarray}
where $\sigma^z$ is the usual Pauli $z$-matrix. Performing the Gaussian integral over the fermions gives the following expression for the SFF
\begin{eqnarray}\label{SFFeff}
\overline{Z(iT) Z(iT)^*} &=&  \prod_{\alpha,\beta} \int \mathscr{D}\Sigma_{\alpha \beta} \mathscr{D}G_{\alpha \beta} e^{N I(G,\Sigma)}\,,
\end{eqnarray}
where
\begin{eqnarray}\label{Icomplex}
I(G,\Sigma) &=& \log\det\left[ - \delta(t-t')\left(\delta_{\alpha \beta}\frac{d}{dt} - i \sigma^z_{\alpha\beta} \frac{a}{T} \right) + \Sigma_{\beta\alpha}(t,t') \right] 
\nonumber\\
&& - \int dt dt' \sum_{\alpha,\beta} \left[ \Sigma_{\beta\alpha}(t',t) G_{\alpha\beta}(t,t') -(-1)^{\alpha+\beta} \frac{m^2}{2} G_{\alpha \beta}(t,t')G_{\beta\alpha}(t',t) \right]. \nonumber\\
\end{eqnarray}

\section{Partition function}\label{appendix:B}

To calculate the partition function (in Euclidean time) for our model, we simply restrict the results (\ref{SFFeff}) and (\ref{Icomplex}) to a single replica, and analytically continue $t\rightarrow -i\tau$, $a\rightarrow ia$ and $m\rightarrow im$, to obtain
\begin{eqnarray}\label{SYK2part}
\overline{Z(\beta)} &=&  \int \mathscr{D}\Sigma \mathscr{D}G e^{NI}, \nonumber\\
I &=& \log\det\left[ - \delta(\tau-\tau')\left(\frac{d}{d\tau} - i\frac{a}{\beta} \right) + \Sigma(\tau,\tau') \right] 
\nonumber\\
&& - \int d\tau d\tau' \left[ \Sigma(\tau',\tau) G(\tau,\tau') + \frac{m^2}{2} G(\tau,\tau')G(\tau',\tau) \right], \label{Ipart}
\end{eqnarray}
with equations of motion given in the large-$N$ saddle point approximation by
\begin{align}\label{partEOM}
    \delta(\tau-\tau'') &= \int d\tau' \left[ - \delta(\tau-\tau') \left(\frac{d}{d\tau} - i \frac{a}{T} \right) + \Sigma(\tau-\tau') \right] G(\tau'-\tau''), \\
    G(\tau) &= -m^2 \Sigma(\tau). \label{B3}
\end{align}
\noindent
We first consider the free theory where $m=0$. The operator $\partial_\tau  - i A(\tau)$ may be explicitly diagonalised as follows \cite{Dunne:1996yb}
\begin{eqnarray}
\left( \frac{d}{d\tau}  - i A(\tau) \right) \psi_n(\tau) & = & \Lambda_n \psi_n(\tau)   \quad \Rightarrow \quad \psi_n(\tau) =  \exp\left\{\int_0^\tau d \tau' ( i  A (\tau')  + \Lambda_n   ) \right\}.\nonumber\\
\end{eqnarray}
Making use of the anti-periodicity of the fermionic field $\psi(0) = -\psi(\beta)$, it follows that 
\begin{eqnarray}
\Lambda_n \beta  + i \int_0^\beta A(\tau) d\tau & = & (2n -1)\pi i \quad \Rightarrow \quad \Lambda_n =  -\frac{i a}{\beta} + \frac{(2n-1)\pi i}{\beta},
\end{eqnarray}
where $n\in\textbf{Z}$ and $a = \int_0^\beta A(\tau) d\tau$ as before. The free partition is then given by
\begin{align}
    Z_0(\beta) &= \left[ \det\left( -\partial_\tau + i A(\tau) \right) \right]^N = \prod_{n=-\infty}^{\infty} \left( \frac{i a}{\beta} -\frac{(2n-1)\pi i}{\beta} \right)^N,
\end{align}
which is formally infinite. We regulate it by dividing by the analogous expression in the absence of the gauge field to obtain
\begin{align}\label{Zfree}
 Z_0(\beta) &= \left[\frac{\det\left( -\partial_\tau +i A(\tau) \right)}{\det\left( -\partial_\tau \right)}\right]^N = \prod_{n=-\infty}^{\infty}  \left[\frac{ -\frac{i a}{\beta} + \frac{(2n-1)\pi i}{\beta}}{\frac{(2n-1)\pi i}{\beta}}\right]^N = \prod_{n=-\infty}^{\infty} \left[ 1 - \frac{a}{(2n-1)\pi}\right]^N  \nonumber\\
    &= \left[\cos\left(\frac{ a}{2}\right)\right]^N. 
\end{align}
This means that the free theory with $m=0$ has a free energy of
\begin{align}
    F_0(\beta) = -\frac{1}{\beta}\log Z_0(\beta) = -\frac{N}{\beta}\log\left[\cos\left(\frac{ a}{2}\right)\right],
\end{align}
which diverges whenever $a$ is an odd multiple of $\pi$. 
\\

\noindent
Generalising to nonzero $m$, direct evaluation of the functional determinant in (\ref{Ipart}) is best avoided in favour of a different method: formulating a differential equation for the free energy \cite{Maldacena:2016hyu, Winer:2020mdc} by exploiting the fact that the only dimensionful quantities in the problem are $m$ and $\beta$, so all $\beta$ dependence must be through $m\beta$. It follows that partial derivatives acting on the free energy are related as $\displaystyle \beta\frac{d}{d\beta}= m\frac{d}{dm}$, so that
\begin{align}\label{a2}
    m^2\beta \frac{dI}{d\beta} &= -2\frac{dI}{d(m^{-2})}  = -\int d\tau d\tau'  \Sigma(\tau'-\tau) \Sigma(\tau-\tau').
\end{align}
Here we have used that the on-shell action (\ref{SYK2part}) only has explicit $m$ dependence since any dependence through the fields doesn't contribute because they satisfy the equations of motion $\frac{\delta I}{\delta \Sigma}=0$. The equations of motion (\ref{partEOM}) may be manipulated to obtain
\begin{align}\label{B18}
    \int d\tau d\tau' \Sigma(\tau-\tau') \Sigma(\tau'-\tau)  = \left[ \beta \frac{d}{d\tau}\Sigma(\tau) -ia \Sigma(\tau) \right]_{\tau\rightarrow0^+} .
\end{align}
Our result for the real-time propagator (\ref{GmSummed}) and (\ref{B3}) gives
\begin{align}
\Sigma(\tau) &= m^2 e^{i \frac{ a}{\beta} \tau} \int_{-2m}^{2m} dy \frac{\sqrt{4 m^2 - y^2}}{2 \pi m^2} \frac{e^{-y \tau}}{ 1 +  e^{- y \beta    } e^{ i a   }   }\,,
\end{align}
which combined with (\ref{a2}) with (\ref{B18}) yields the desired differential equation
\begin{align}
    \frac{dI}{d\beta}
    &= \int_{-2m}^{2m} dy \frac{  y \sqrt{4 m^2-y^2}}{2 \pi m^2 ( 1+ e^{i a-\beta y}) }.
\end{align}
Integrating w.r.t. $\beta$ (swapping the order of integration) yields
\begin{align}
    I &= \int_{-2m}^{2m} dy \frac{\sqrt{4 m^2-y^2} \log \left(e^{\beta  y}+e^{i a}\right)}{2 \pi  m^2} + C.
\end{align}
where $C$ is a constant of integration. The integrand may be expanded as a polynomial in $\beta$ using
\begin{align}
    \sqrt{4 m^2-y^2} \log \left(e^{\beta  y}+e^{i a}\right) &=  \sum_{n=0}^\infty y^n\sqrt{4m^2-y^2} \frac{f_n(e^{ia})}{n!(1+e^{ia})^n}  \beta^n
\end{align}
where $f_n$ is a polynomial, given for the first few terms by
\begin{align*}
    f_0(x) &= \log\left(1+x\right)\\
    f_1(x) &= 1,\\
    f_2(x) &= x,\\
    f_3(x) &= x^2-x,\\
    f_4(x) &= x^3-4x^2+x,\\    
    f_5(x) &= x^4 -11x^3+11x^2-x.
\end{align*}
We may then perform the integral term by term, to obtain
\begin{align}
    I &= - \frac{1}{\sqrt{\pi }} \sum_{n=0}^\infty \frac{2^n}{n!}  \frac{ \Gamma \left(\frac{n+1}{2}\right)}{\Gamma \left(\frac{n}{2}+2\right)} \frac{f_n(e^{ia})}{(1+e^{ia})^n}  m^n\beta^n \nonumber\\ 
    &= \log\cos\left(\frac{a}{2}\right) + \frac{1}{8} \sec ^2\left(\frac{a}{2}\right) m^2\beta^2 + \mathcal{O}( m^4 \beta^4) ,
\end{align}
where we have fixed the integration constant $C = \log(e^{-ia/2}/2)$ to ensure that we recover (\ref{Zfree}) in the limit $m\rightarrow0$. Some remarks are in order. Note that this expression is invariant under large gauge transformations. It diverges when $a$ is equal to an odd multiple of $\pi$. We verify that it is equal, order by order, to the exponent of the $mT\ll0$ SFF (\ref{order8}) up to normalisation (unlike the partition function, the SFF must satisfy $g(0)=1$), the replacement $m\rightarrow im$ (to account for the Wick rotation to real time) and an overall factor of $2$ (since the SFF is the square of a partition function). It therefore also serves as a useful check of the early time SFF result in the main text, which is given by the disconnected component of the SFF (\ref{gd}) as expected \cite{Cotler:2016fpe}.

\section{SFF fluctuations: Mass matrices}

\label{MassMatrixAppendix}

For each frequency, only saddle point solutions whose mass matrices have positive real parts (for both replica indices) will contribute in the large $N$ saddle point approximation. In this appendix we carefully check this condition for each saddle point in each of the four regions.  \\

\textbf{Region $A$:} $|x_n\pm b|\leq2$, from which it follows that $|b|<2$ and $|x_n|<2-|b|$. Note that for $|b|>2$, this saddle point $A$ does not contribute at all. The solution in this region is given by 
\begin{align}\label{Asad}
\mathcal{S}_{\alpha\beta}^{nn'(A)} =\frac{1}{2} \left[ -i(\xi_\alpha x_n+b) + c_\alpha^n \sqrt{4-(\xi_\alpha x_n +b)^2}\right]\delta^{nn'}_{\alpha\beta},
\end{align}
We can independently choose $c_1^n$ and $c_2^n$, giving four possible solutions for each $n$. The corresponding mass matrices are given by 
\begin{align}
    M^{nn'(A)}_{\alpha\beta} = 1 + \frac{1}{4} \left[ -i(\xi_\alpha x_n+b) + c_\alpha^n \sqrt{4-(\xi_\alpha x_n +b)^2}\right]
    \left[ -i(\xi_\beta x_{n'}+b) + c_\beta^{n'} \sqrt{4-(\xi_\beta x_{n'} +b)^2}\right]
\end{align}
where $|\xi_\alpha x_n+b|\leq2$ and $|\xi_\beta x_{n'}+b|\leq2$. It can be verified that Re$\left[M^{nn'}_{\alpha\beta}\right]\geq0$ for all choices of $c_\alpha^{n}$ and $c_\beta^{n'}$. We note that when $c_\alpha^{n}=-c_\beta^{n'}$ and $\xi_\alpha x_n=\xi_\beta x_{n'}$ ($\alpha\neq\beta$ and $n=-n'$)\footnote{The other case $\alpha=\beta$, $n=n'$ which satisfies $\xi_\alpha x_n=\xi_\beta x_{n'}$ is ruled out by the condition $c_\alpha^{n}=-c_\beta^{n'}$.} the corresponding mass matrices vanish: $M^{n,-n}_{12}=M^{n,-n}_{21}=0$. This signals the presence of zero modes, whose significance is discussed in the main text. Since the mass matrix is always non-negative, all four saddle point solutions contribute to the SFF in this region. Note also that the mass matrices for the two saddles with $c_1=c_2$ are complex conjugates of each other
\begin{align}\label{AMc}
    \left( M^{nn'(A)}_{\alpha\beta}\big|_{c_1^{n}=c_2^{n'}=\pm1} \right)^* = M^{nn'(A)}_{\alpha\beta}\big|_{c_1^{n}=c_2^{n'}=\mp1},
\end{align}
a fact that will become relevant when evaluating the late time SFF later.
\\ \\
\textbf{Region $D$:} $|x_n\pm b|>2$. In this case the solution is completely determined and there is only one choice for each frequency. This regime can be written $|x_n|>2+|b|$ and $|x_n|<|b|-2$ (when $|b|<2$). The solution here is given by 
\begin{align}\label{Dsad}
\mathcal{S}_{\alpha\beta}^{nn'(D)} =\frac{1}{2} \left[ -i(\xi_\alpha x_n +b) + i\text{sgn}(\xi_\alpha x_n +b) \sqrt{(\xi_\alpha x_n +b)^2-4}\right]\delta^{nn'}_{\alpha\beta},
\end{align}
so we have 
\begin{align}\label{DM}
    M^{nn'(D)}_{\alpha\beta} =& 1 + \frac{1}{4} \left[ -i(\xi_\alpha x_n +b) + i\text{sgn}(\xi_\alpha x_n +b) \sqrt{(\xi_\alpha x_n +b)^2-4}\right] \nonumber\\
    &\qquad \times \left[ -i(\xi_\beta x_{n'} +b) + i\text{sgn}(\xi_\beta x_{n'} +b) \sqrt{(\xi_\beta x_{n'} +b)^2-4}\right]
\end{align}
where $|\xi_\alpha x_n+b|>2$ and $|\xi_\beta x_{n'}+b|>2$. In this case Re$\left[M^{nn'}_{\alpha\beta}\right]= M^{nn'}_{\alpha\beta}>0$ for all allowed frequencies and replica indices. Of course since $M^{nn'}_{\alpha\beta}$ is real and independent of the constants $c_\alpha^n$, we have that 
\begin{align}\label{DMc}
    \left( M^{nn'(D)}_{\alpha\beta}\big|_{c_1^{n}=c_2^{n'}=\pm1} \right)^* = M^{nn'(D)}_{\alpha\beta}\big|_{c_1^{n}=c_2^{n'}=\mp1}.    
\end{align}

\textbf{Region $B$:} $|x_n+b|\leq2$, $|x_n-b|>2$. We can choose the single constant $c_{1}^n$, giving two possible solutions for each $n$. We can write the region as
\begin{align}
    |b|>2:& \qquad |x_n+b|<2, \\
    |b|\leq 2:& \qquad |x_n+ \text{sgn}(b) 2|< |b|.
\end{align}
The solutions here is given by 
\begin{align}
\mathcal{S}_{\alpha\beta}^{nn'(B)} = & 
=\frac{1}{2} \left[ -i(\xi_\alpha x_n+b) + c_\alpha^n \sqrt{4-(\xi_\alpha x_n +b)^2}\right]\delta^{nn'}_{\alpha1} \nonumber\\
& +\frac{1}{2} \left[ -i(\xi_\beta x_{n} +b) + i\text{sgn}(\xi_\beta x_{n} +b) \sqrt{(\xi_\beta x_{n} +b)^2-4}\right]\delta^{nn'}_{2\beta}.
\end{align}
The first diagonal component of the mass matrix is given by
\begin{align}
    M^{nn'(B)}_{11} =&  M^{nn'(A)}_{11},
\end{align}
where $|x_n+ b|\leq2$ and $|x_{n'}+ b|\leq2$. It follows that Re$\left[M^{nn'}_{11}\right]>0$. Note that the zero modes from $A$ do note feature here since they only exist for $\alpha\neq\beta$. The second diagonal component is given by
\begin{align}
    M^{nn'(B)}_{22} =&  M^{nn'(D)}_{22},
\end{align}
where $|x_n- b|>2$ and $|x_{n'}- b|>2$. In this case Re$\left[M^{nn'}_{22}\right]>0$ as above. Finally, we have the replica non-diagonal components given by
\begin{align}
    M^{nn'(B)}_{\alpha\neq\beta} =& 1 + \frac{1}{4} \left[ -i(\xi_\alpha x_n +b) + i\text{sgn}(\xi_\alpha x_n +b) \sqrt{(\xi_\alpha x_n +b)^2-4}\right] \nonumber\\
    &\qquad \times \left[ -i(\xi_\beta x_{n'} +b) + c_\beta^{n'} \sqrt{(\xi_\beta x_{n'} +b)^2-4}\right],
\end{align}
where $|\xi_\alpha x_n+b|\leq2$ and $|\xi_\beta x_{n'}+b|>2$. In all of the allowed frequency regimes, we have that Re$\left[M^{nn'}_{\alpha\neq\beta}\right]>0$. Thus the saddles for both choices of $c_1^n$ contribute at each frequency in this regime. In all three cases we have that
\begin{align}\label{BMc}
    \left( M^{nn'(B)}_{\alpha\beta}\big|_{c_1^{n}=c_2^{n'}=\pm1} \right)^* = M^{nn'(B)}_{\alpha\beta}\big|_{c_1^{n}=c_2^{n'}=\mp1}.
\end{align}

\textbf{Region $C$:} $|x_n-b|\leq2$, $|x_n+b|>2$. We can choose the single constant $c_2^n$, giving two possible solutions for each $n$. The region can be written as
\begin{align}
    |b|>2:& \qquad |x_n-b|<2, \\
    |b|\leq2:& \qquad |x_n - \text{sgn}(b) 2|<|b|.
\end{align}
Similarly to the above case, we find that the saddles for both choices of $c_2^n$ contribute at each allowed frequency, since
\begin{align}
    M^{nn'(C)}_{11} &=  M^{nn'(D)}_{11} >0, \\
    M^{nn'(C)}_{22} &=  M^{nn'(A)}_{22} >0, \\
    M^{nn'(B)}_{\alpha\neq\beta} &> 0.
\end{align} 
Also,
\begin{align}\label{CMc}
    \left( M^{nn'(C)}_{\alpha\beta}\big|_{c_1^{n},c_2^{n'}=\pm1} \right)^* = M^{nn'(C)}_{\alpha\beta}\big|_{c_1^{n},c_2^{n'}=\mp1}.
\end{align}
The above constitutes an extensive list of all mass matrices one may encounter when performing the saddle point approximation for the SFF.  

\begin{figure}
\centering
    \begin{subfigure}{}
    \centering
    \includegraphics[scale=0.3]{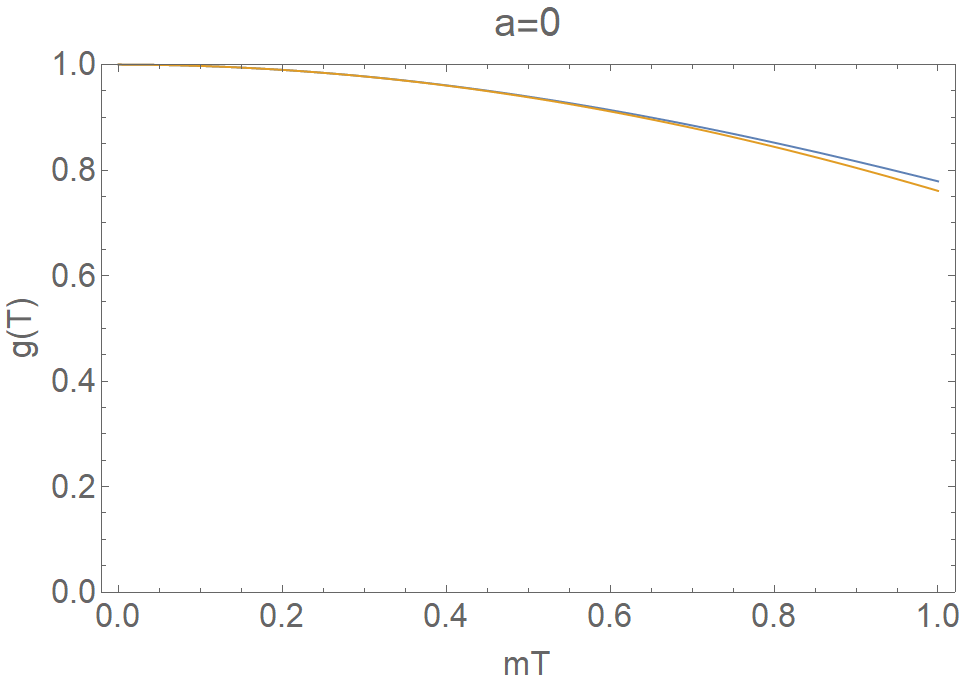} 
    \end{subfigure}
    \begin{subfigure}{}
    \centering
    \includegraphics[scale=0.3]{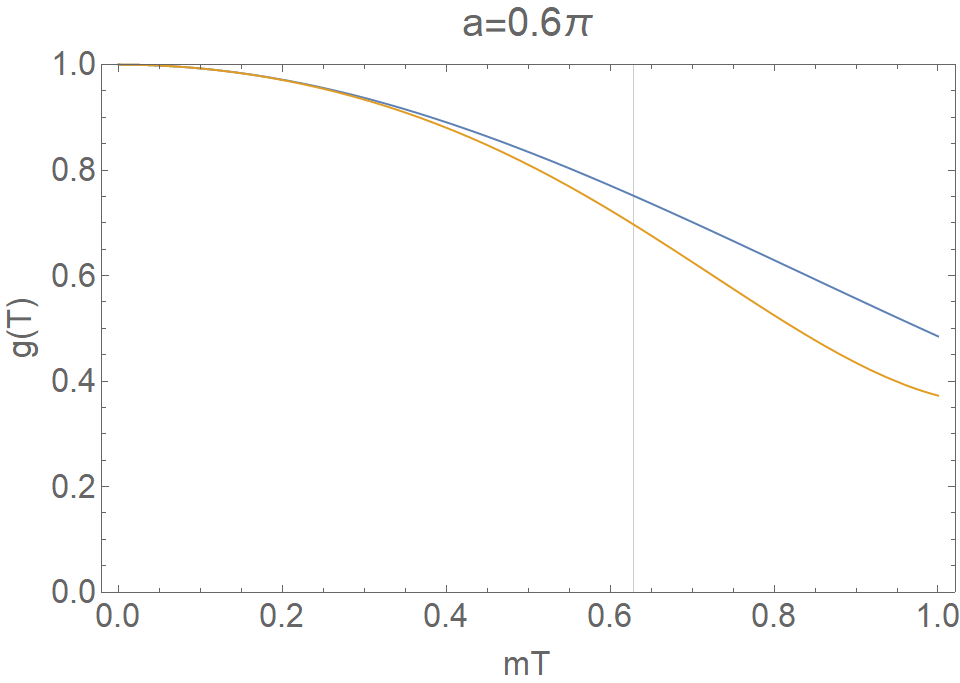} 
    \end{subfigure}
    \begin{subfigure}{}
    \centering
    \includegraphics[scale=0.3]{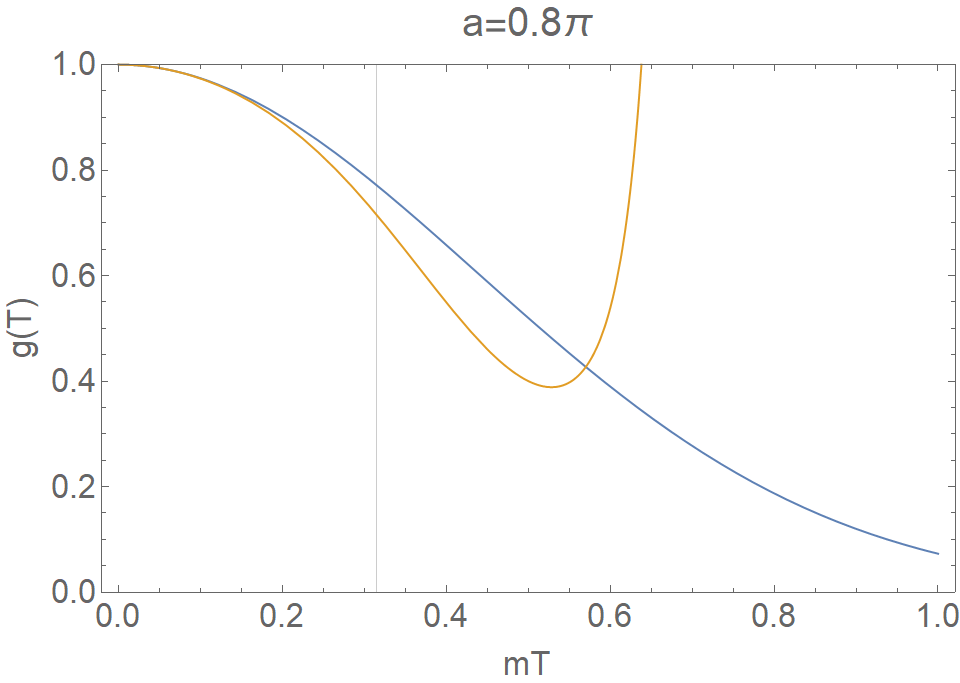} 
    \end{subfigure}
    \begin{subfigure}{}
    \centering
    \includegraphics[scale=0.3]{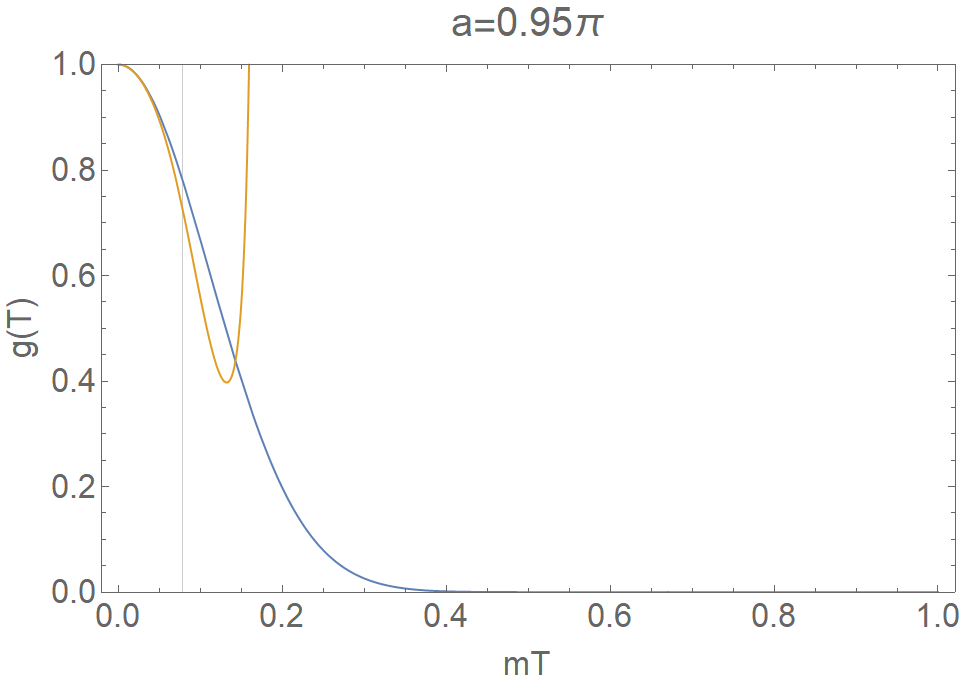} 
    \end{subfigure}
    \caption{SFF (\ref{order8}) for $mT\ll1$ for various $a$, to $\mathcal{O}(m^2T^2)$ (blue) and order $\mathcal{O}(m^8T^8)$ (yellow) in the exponent. Grey vertical lines mark $mT=\frac{\pi-a}{2}$. Here we have set $N=1$, for large $N$ the decay is much more rapid.}
    \label{smallmTplot}
\end{figure}

\section{$mT\ll1$ SFF}\label{appendix:C}

\begin{align}\label{order8}
    g(T) =& \exp \Bigg\{N\Bigg(-\frac{1}{4}m^2 T^2 \sec ^2\left(\frac{a}{2}\right) + \frac{1}{48} m^4 T^4 (\cos a-2) \sec ^4\left(\frac{a}{2}\right) \nonumber\\
    &\qquad-\frac{1}{2304} m^6 T^6 (-26 \cos a+\cos2a+33) \sec ^6\left(\frac{a}{2}\right) \nonumber\\
    &\qquad -\frac{m^8 T^8 (1191 \cos a-120 \cos (2 a)+\cos (3 a)-1208) \sec ^8\left(\frac{a}{2}\right)}{184320} + \mathcal{O}(m^{10}T^{10}) \Bigg) \Bigg\}.
\end{align}
Note that this expression is invariant under large gauge transformations $a\rightarrow a+ 2k\pi$, $k\in \textbf{Z}$. As $a$ approaches $\pi$, our result becomes trustworthy only for smaller and smaller times $mT<\frac{\pi-a}{2}$, since we have assumed this in order to neglect the contribution from the saddle $C$. For $mT<\frac{\pi-a}{2}$ we always see the characteristic slope of the SFF, see figure \ref{smallmTplot}.

\section{SFF for $\pi-a<2mT\ll1$}\label{appendix:D}

That gauge field configurations for which $a=\int_0^\beta dt A(t)$ is an odd multiple of $\pi$ are unphysical can be motivated by the fact that in such cases, we are effectively shifting all fermionic Matsubara frequencies to bosonic Matsubara frequencies, for which $\omega_n=\frac{2n\pi}{T}$ for integer $n$. This in turn violates the antiperiodicity requirement on our fields across the thermal circle. Indeed, the partition function diverges in such cases, as is shown in Appendix (\ref{appendix:B}).
\\

\begin{figure}
\centering
\begin{subfigure}{}
    \centering
    \includegraphics[scale=0.3]{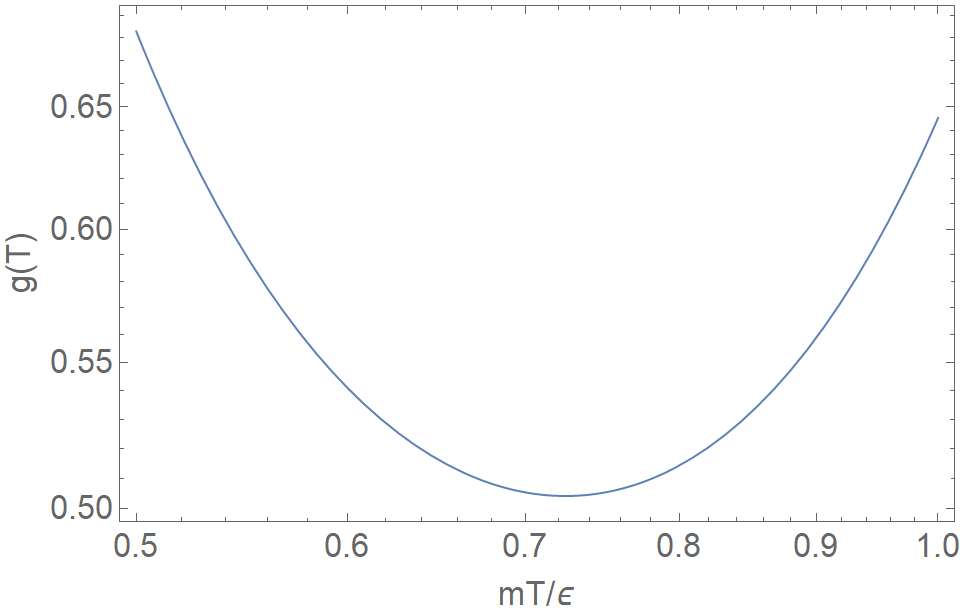} 
    \end{subfigure}
    \begin{subfigure}{}
    \centering
    \includegraphics[scale=0.3]{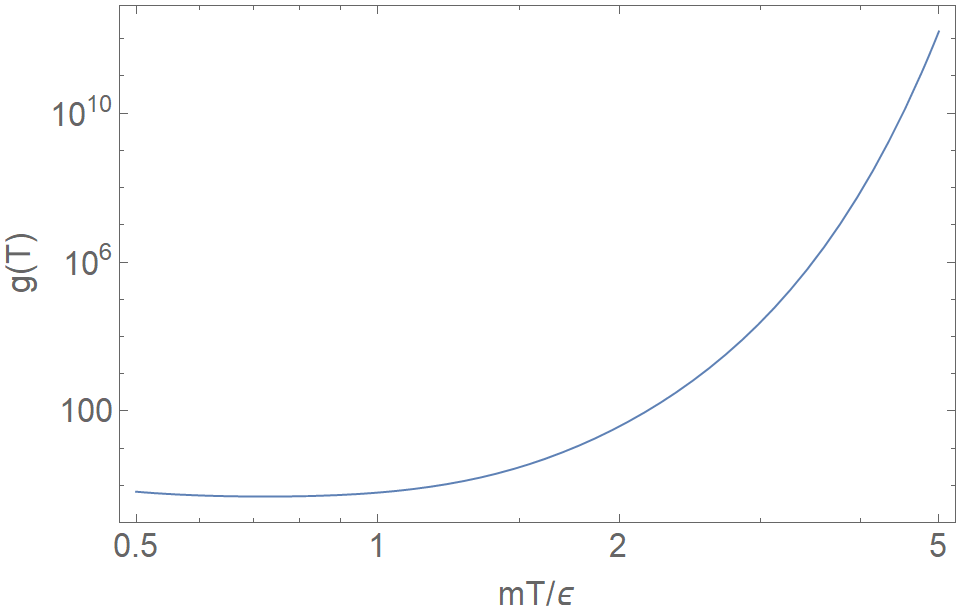} 
    \end{subfigure}
    \caption{Log-log plots of the SFF (\ref{mTll1SFFa}) for $a=\pi-\epsilon$ and $0<\epsilon<2mT\ll1$, with $N=1$.}
    \label{epsplot}
\end{figure}

In evaluating the early time SFF (\ref{SFFbg0Ta}), as $a$ approaches $\pi$ we are eventually forced to take the contribution from frequencies $x_n$ in region $C$, where $|n\pi-a|<2mT$, into account. To this end, we take $\epsilon \equiv \pi-a > 0$ where $0<\epsilon<2mT\ll1$, so that the $n=1$ frequency contribution now comes from region $C$. We then evaluate the SFF (\ref{SFFbg0Ta}) the same as before, except that the $n=1$ contribution now comes from region $C$ and not region $D$
\begin{align}\label{mTll1SFFa}
    \overline{Z(iT)Z^*(iT)} = & \sum_{c, \tilde{c} =\pm1} \exp\Bigg\{ N \left[ I_{C}^{(c)}\left(\frac{\pi}{mT}\right) + I_{C}^{(\tilde{c})}\left(\frac{\pi}{mT}\right) + 2 \sum_{n>1,\text{odd}} I_D(x_n)  \right] \Bigg\}.
\end{align}
Expanding to $\mathcal{O}(m^2T^2)$, performing the sums and implementing the normalisation (including dividing by the factor of 4 since we sum over 4 saddles) we obtain
\begin{align}\label{mTll1SFFa}
    g(T) = & \frac{1}{4} \left( 1 + \frac{m^2 T^2}{e \epsilon^2} \exp\left\{\frac{m^2 T^2}{\epsilon ^2}+\frac{\epsilon ^2}{2 m^2 T^2}\right\} \right)^{2N} \exp\left\{N\frac{m^2 T^2}{2 \cos\epsilon -2}\right\} .
\end{align}
This result is valid to all orders in $\epsilon$ for times $\epsilon<2mT\ll1$ (indeed it diverges as $mT\rightarrow 0$ for any finite $\epsilon$). We verify that the same result is obtained had we set $a=3\pi-\epsilon$ at the start of the calculation and accordingly evaluated the $C$ saddles at $x_n=3\pi/mT$. We expect that the result is invariant under all large gauge transformations $a\rightarrow a+2\pi k$, $k\in Z$ in this sense.
\\

We plot our result (approximating $\cos\epsilon\approx1-\epsilon^2/2$ to obtain a result which is a function only of the combination $mT/\epsilon$) in figure \ref{epsplot}. On the left we zoom into the region $\epsilon<2mT<2\epsilon$ and we see the end of the slope and the beginning of a region of increasing behaviour: this is the characteristic ``dip'' of the SFF. We may numerically estimate the dip time as
\begin{align}\label{diptimeest}
    m T_\text{dip} \approx 0.73(\pi-a),
\end{align}
valid for values of the the gauge field strength $a$ such that $0<\pi-a \ll 1$. If we simply ignore this range of validity and set $a=0$, we obtain a dip time (for $m=1$) for the SYK$_2$ model of 
\begin{align}
    T_\text{dip} \approx 0.73\pi. 
\end{align}
Using equations (17) and (33) of \cite{Winer:2020mdc} for the SFF\footnote{Note that we are here comparing the dip times of the complex and Majorana SYK$_2$ models. This is justified since we expect these results to differ only by a power of 2 (accounting for the difference between the number of degrees of freedom in the models) which does not affect the dip time.}, we obtain a dip time of 
\begin{align}
    T_\text{dip} \approx 0.76\pi. 
\end{align}
Using equation (5b) of \cite{Liao:2020lac} for the SFF, with the chemical potential switched off, we obtain a dip time of \begin{align}
    T_\text{dip} \approx 0.8\pi. 
\end{align}
It thus appears that our result for the $0<\pi-a \ll 1$ dip time remains valid for values of $a$ outside of these bounds. It would be interesting to investigate the reasons for this further, which we leave to future work.
\\

Note that the increase we see after the dip time is exponential in $N$, whereas we know that the late time ramp is exponential in $\log N$. This exponential in $N$ behaviour therefore likely corresponds to fluctuations\footnote{See figure 1 of \cite{Winer:2020mdc} or \cite{Liao:2020lac}.} which are known to exist at the start of the SYK$_2$ exponential ramp, and cannot be trusted up to the values of $mT$ plotted on the right of figure \ref{epsplot}.

\section{Vanishing of the $mT\gg1$ sum over saddles}\label{appendix:F}

We may use the following features of the real and imaginary parts of the on-shell actions 
\begin{align}
    [I_A^{(1,1)}(x)]^* &= I_A^{(-1,-1)}(x), \qquad [I_A^{(1,-1)}(x)]^* = I_A^{(-1,1)}(x) \nonumber\\
    \text{Re}[I_A^{(1,1)}(x)] &= \text{Re}[I_A^{(1,-1)}(x)], \qquad [I_C^{(1)}(x)]^* = I_C^{(-1)}(x),
\end{align}
to perform the sum over saddles $\{c\},\{\tilde{c}\}$ in (\ref{SFFbl0Ta}) and rewrite the SFF valid for large times $2mT>|a|$ as
\begin{eqnarray}\label{SFFbl0T}
   &&\overline{Z(iT)Z^*(iT)} \nonumber\\
   &=&  \prod_{\substack
  {n\pi>0\\n \ \text{odd}}}^{2mT-|a|} \left[ e^{2NI_A^{(1,1)}(x_n)} + e^{2N [I_A^{(1,1)}(x_n)]^*} + e^{2NI_A^{(1,-1)}(x_n)} + e^{2N [I_A^{(1,-1)}(x_n)]^*} \right] \nonumber\\
   && \prod_{\substack{ n\pi> 2mT-|a|\\n \ \text{odd}}}^{2mT+|a|} \left[ e^{2N I_C^{(1)}(\text{sgn}(a) x_n)} + e^{2N [I_C^{(1)}(\text{sgn}(a) x_n)]^*} \right] \prod_{\substack
  { n\pi> 2mT+|a|\\n \ \text{odd}}}^{\infty} e^{ 2N I_D(x_n)} \nonumber\\
  &=&  \prod_{\substack
  {n\pi>0\\n \ \text{odd}}}^{2mT-|a|} 2 e^{2N\text{Re}[I_A^{(1,1)}(x_n)]} \left[
  \cos \left( 2N \text{Im}[I_A^{(1,1)}(x_n)]\right) +  \cos\left(2N \text{Im}[I_A^{(1,-1)}(x_n)]\right) \right] \nonumber\\
   && \prod_{\substack{ n\pi> 2mT-|a|\\n \ \text{odd}}}^{2mT+|a|} 4 \cos^2\left(N \text{Im}[I_C^{(1)}(\text{sgn}(a) x_n)]\right) e^{2N\text{Re}[I_C^{(1)}(\text{sgn}(a) x_n)]} \prod_{\substack
  { n\pi> 2mT+|a|\\n \ \text{odd}}}^{\infty} e^{ 2N I_D(x_n)}. \nonumber\\
\end{eqnarray}
In the limit $mT\gg1$, the variable $x_n = \frac{n\pi}{mT}$ becomes continuous and the late time expression for the SFF is proportional to
\begin{eqnarray}
   &&\prod_{x=0}^{2-|b|} \left[
  \cos \left( 2N \text{Im}[I_A^{(1,1)}(x)]\right) +  \cos\left(2N \text{Im}[I_A^{(1,-1)}(x)]\right) \right] \nonumber\\
  &=& \prod_{x=0}^{2-|b|} 
  \cos \left[ N \left(\text{Im}[I_A^{(1,1)}(x)+I_A^{(1,-1)}(x_n)]\right) \right]  \cos \left[ N\left(\text{Im}[I_A^{(1,1)}(x_n)-I_A^{(1,-1)}(x)]\right) \right] \nonumber\\
\end{eqnarray}
This factor, and so the SFF, vanishes when the argument of either cosine is equal to $k\pi/2$, $k$ odd. For $k$ such that $k=\pm2N$ this condition becomes
\begin{align}
    \text{Im}[I_A^{(1,1)}(x) \pm I_A^{(1,-1)}(x)] = \pm\pi.
\end{align}
It is easily verified numerically that this condition is always satisfied for some $x\in[0,2-|b|)$ when $|b|\leq1$. The SFF (\ref{SFFbl0T}) is also proportional to
\begin{align}
    \prod_{x= 2-|b|}^{2+|b|} \cos^2\left(N \text{Im}[I_C^{(1)}(\text{sgn}(b)x)]\right),
\end{align}
and it can similarly be verified that this factor vanishes for some $x\in(2-|b|,2+|b|)$ when $|b|>1$. Thus the SFF obtained purely from the saddle point contributions vanishes at late times for any $b=\frac{a}{mT}$.

\end{document}